\documentclass[amsmath, preprintnumbers, 10pt, twocolumn, pre bibliograpy, floatfix]{revtex4-1}

\usepackage{graphicx}
\usepackage{color}
\usepackage{amsmath}
\usepackage{amssymb}

\usepackage{physics}
\usepackage{float}
\usepackage{caption}
\usepackage[font=small,labelfont=bf,justification=RaggedRight,format=plain, singlelinecheck=false]{caption}
\usepackage{subcaption}
\usepackage{array}
\usepackage{hyperref}
\usepackage{cleveref}
\usepackage{amsmath}
\usepackage{stackengine}
\usepackage{mathrsfs}
\renewcommand{\theequation}{\arabic{equation}}
\numberwithin{equation}{section}
\usepackage{xr}
\makeatletter

\begin{document}
\title{A mechanism for the generation of robust circadian oscillations through ultransensitivity and differential binding affinity} 
\author{Agnish Kumar Behera and Clara del Junco and  Suriyanarayanan Vaikuntanathan}
\affiliation{Department of Chemistry and The James Franck Institute, University of Chicago, Chicago, IL, 60637}
\begin{abstract}
Biochemical circadian rhythm oscillations play an important role in many signalling mechanisms. In this work, we explore some of the biophysical mechanisms responsible for sustaining robust oscillations by constructing a minimal but analytically tractable model of the circadian oscillations in the KaiABC protein system found in the cyanobacteria \textit{S. elongatus}. In particular, our minimal model explicitly accounts for two experimentally characterized biophysical features of the KaiABC protein system, namely, a differential binding affinity and an ultrasensitive response. Our analytical work shows how these mechanisms might be crucial for promoting robust oscillations even in sub optimal nutrient conditions. Our analytical and numerical work also identifies mechanisms by which biological clocks can stably maintain a constant time period under a variety of nutrient conditions. Finally, our work also explores the thermodynamic costs associated with the generation of robust sustained oscillations and shows that the net rate of entropy production alone might not be a good figure of merit to asses the quality of oscillations. 
\end{abstract}
\maketitle 

\section{Introduction}
\renewcommand*{\thesection}{\arabic{section}}
	
Most living organisms, ranging from simple single celled organisms like cyanobacteria to multicellular organisms possess an internal clock which is entrained with the day-night cycle \cite{Mohawk2012, Kondo2000, BLAU2001287, COLLINS2006348, Jeanne2005}. The fidelity and robustness of this clock is crucial for the well being and survival of the organism \cite{Dubowy2017, MartinsE11415, Ouyang8660, Liaoe2022516118}. The time period of the internal clock has, for example, been found to be robust with respect to changes in the temperature, nutrient conditions, and pH \cite{Phong2013, Clodong2007, AVELLO2021110495, DOVZHENOK20151830}. Understanding the biochemical and thermodynamic underpinnings of such robust behavior remains an important challenge given the crucial biological role of the internal clock. 

\begin{figure}[thpb]
    \centering
    \includegraphics[width = 0.35\textwidth]{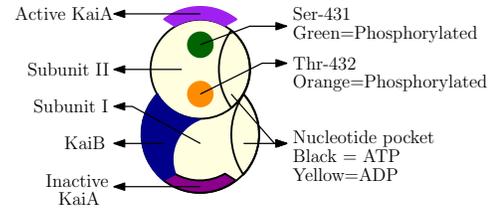}
    \caption{KaiC monomer. The KaiC protein exists as a hexamer and each monomer consists of 2 domains, CI and CII. The CII domain has two phosphorylation sites, Ser-431 and Thr-432, a KaiA binding site and a nucleotide binding site (which binds either ATP or ADP). The CI domain binds to KaiB and helps sequester KaiA. Subsequently, the KaiABC complex will be denoted using $^{-/A/B}CI_{TP/DP}-^{-/A}CII_{TP/DP}^{U/T/S/D}$. Here TP/DP denotes ATP/ADP attached to the domain. U denotes none of the sites in CII are phosphorylated, S means only Serine site is phosphorylated, T means Threonine site is phosphorylated, D denotes doubly phosphorylated form. A attached to CI denotes sequestered KaiA, A attached to CII denotes active KaiA acting as an assistant in phosphorylation. B attached to CI implies inactive form which will start sequestering KaiA.}
    \label{KaiCMonomer}
\end{figure}

The KaiABC protein system (see Fig.~\ref{KaiCMonomer}) provides a minimal biochemically tractable model to explore the above mentioned questions. The KaiABC system is found in cyanobacteria \textit{S. elongatus} where it plays the role of regulating the circadian cycle. The KaiABC system consists of three proteins, KaiA, KaiB and KaiC \cite{Paijmans2017}. In vitro, the system of KaiABC proteins undergoes sustained oscillations as evidenced by the phosphorylation state of the KaiC protein. These oscillations have been shown to have many of the same robust features as those observed in the circadian oscillations they support in cyanobacteria \cite{Rust220, Tomita251}. The KaiABC model system has been probed in many experimental and theoretical studies\cite{Phong2013,Paijmans2017}. These have elucidated some of the necessary requirements for the generation of sustained oscillations \cite{Phong2013, Hong2020, Hatakeyama2015, Nishiwaki13927, Rust16760, Rust220,Zhang2020}. Despite these advances, understanding the biochemical and biophysical driving forces that are responsible for sustaining robust oscillations remains an open question \cite{Zhang2020, Phong2013, Paijmans2017, Tomita251, Nishiwaki13927, NISHIWAKI201218030}. 

In this paper, we build on recent experimental and modelling work in Ref~\cite{Hong2020} and show how a particular ultrasensitive switch in the KaiABC biochemical circuit can control the quality and robustness of oscillations. In particular, in Ref~\cite{Hong2020}, the authors identify a previously underappreciated ultrasensitive response in the phosphorylation levels of the KaiC proteins as the concentration of the KaiA proteins is tuned. The KaiB proteins play no role in this ultrasensitive response. It was postulated in Ref~\cite{Hong2020} that this ultrasensitive switch plays a central role in ensuring robust oscillations. Specifically, the ultrasensitive switch allows the system to exhibit sustained oscillations even at low levels of the energy rich molecule, ATP~\cite{Hong2020}. Motivated by this work, we build a minimal Markov state model that provides analytical insight for how an ultrasensitive KaiA-KaiC switch can modulate the quality of oscillations. Our minimal model also allows us to analytically study how another biophysical driving force, namely the differential affinity of the different forms of KaiC to KaiA~\cite{Phong2013, Nishiwaki13927, NISHIWAKI201218030, Rust220} also control oscillations. Finally, our minimal Markov state model allows us to comment on the thermodynamic costs associated with setting up robust oscillations in the KaiABC system. 

The rest of the paper is organized as follows. We first briefly review the salient features of the KaiABC biochemical circuit and then outline our minimal model. This model captures the above mentioned features of the KaiABC circuit, namely the differential affinity of KaiC to KaiA binding, and the ultrasensitive response of KaiC phosphorylation levels to changes in KaiA concentration. It also additionally accounts for many other experimentally characterized biophysical forces~\cite{Paijmans2017}. We then write down a stochastic master equation to describe the dynamics of our model. This stochastic master equation is non-linear in the probability. The non-linearity is due to the various feedback mechanisms that are necessary for sustaining oscillations. Interestingly, by solving the non-linear stochastic master equation, we are able to analytically describe the emergence of global oscillations in response to changing the differential affinity \cite{Zhang2020}. Our model allows us to obtain approximate analytical solutions that provide qualitative insight into how tuning ultrasensitivity tunes the quality of oscillations. Crucially, our results allow us to elucidate how an ultrasensitive switch can support oscillations even at a lower concentration of ATP. Our results also allow us to explain how the time period of oscillations can be robustly maintained even as the concentration of ATP is tuned, a phenomenon known as affinity compensation. Finally, we comment on the thermodynamic costs associated with sustaining robust oscillations.

\renewcommand*{\thesection}{\Roman{section}}
\section{KaiABC Oscillator and Model details}
\renewcommand*{\thesection}{\arabic{section}}

\label{modelsec}

\begin{widetext}

\begin{figure}[thb]
    \begin{subfigure}[b]{0.25\textwidth}
    \centering
    \includegraphics[width = \textwidth]{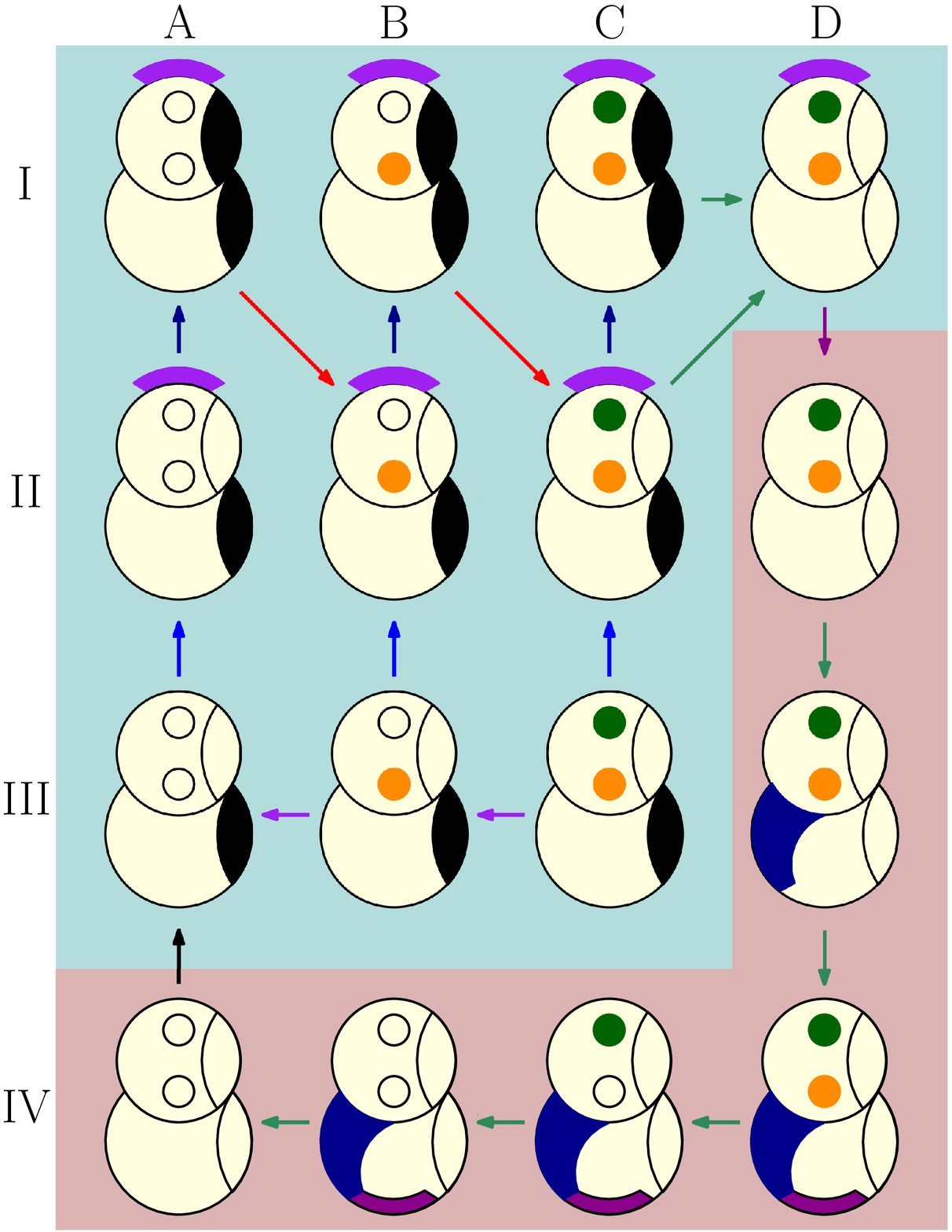}
    \caption{One complete oscillatory cycle of a KaiC monomer.}
    \label{MonomerCycle}
    \end{subfigure}
    \hfill
    \begin{subfigure}[b]{0.7\textwidth}
    \centering
    \includegraphics[width = \textwidth]{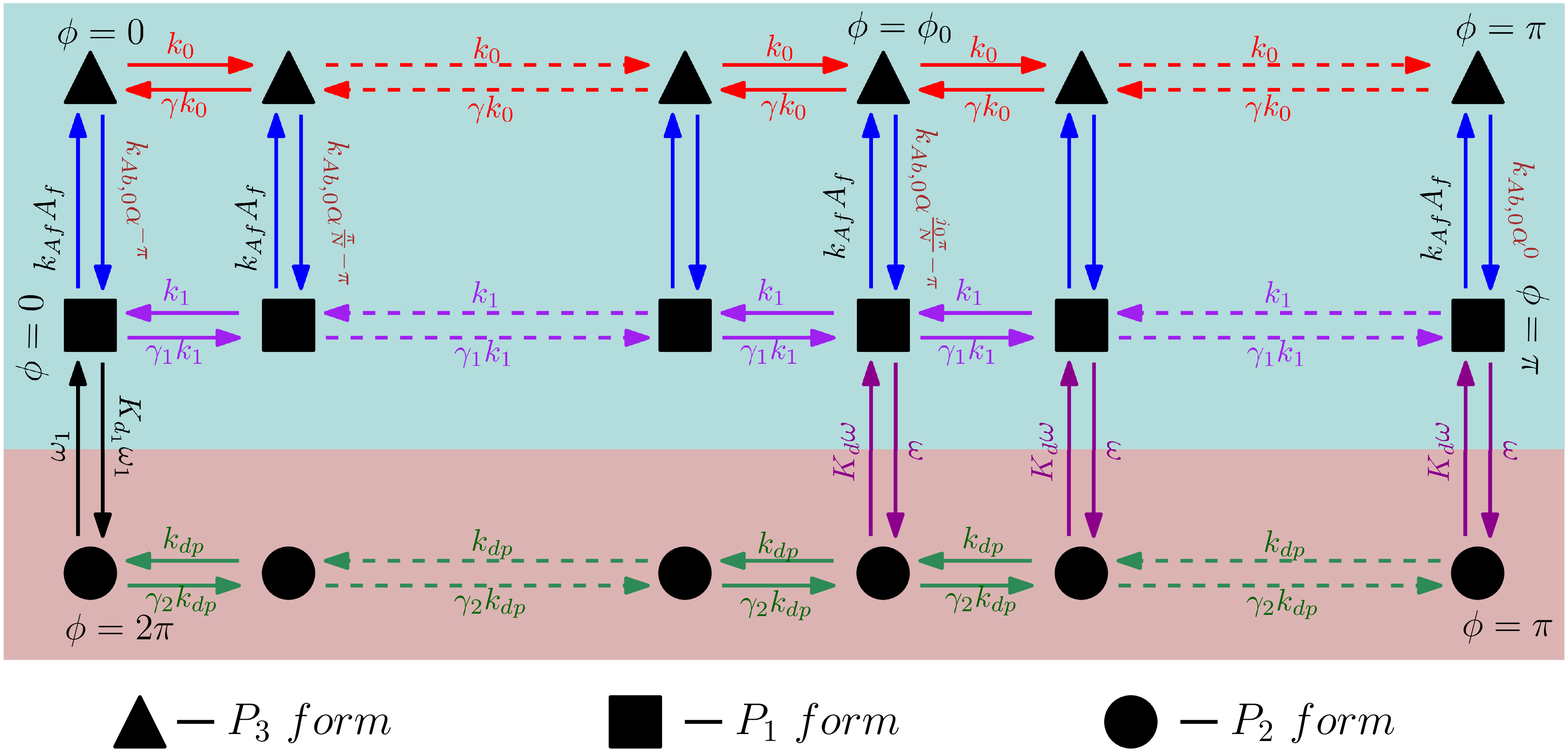}
    \caption{Minimal coarse grained model of KaiABC oscillator.}
    \label{Model}
    \end{subfigure}
    \caption{In Fig. \ref{MonomerCycle}, rows are labelled by I, II, III, IV and columns are labelled by A, B, C, D. In Fig. \ref{MonomerCycle}, the color in the reaction arrows correspond to those in Fig. \ref{Model}. Active conformations are denoted using a \textit{cyan} background and inactive conformations are denoted using a \textit{red} background. In our model (Fig. \ref{Model}), the horizontal axis represents the amount of phosphorylation in the system, with $\phi = 0$ and $\phi = 2\pi$ corresponding to the completely dephosphorylated state and $\phi = \pi$ corresponding to the completely phosphorylated hexamer. The phosphorylation function is a linearly increasing function, 0 at $\phi = 0$, 1 at $\phi = \pi$ and then symmetrically decreasing from $\phi = \pi$ to $2\pi$. Thus, phosphorylation, $\mathscr{P} = \sum_{\phi} \frac{\phi}{\pi} (P_1(\phi) + P_3(\phi)) + \sum_{\phi}(2-\frac{\phi}{\pi})P_2(\phi)$. Changes in the phosphorylation levels of the KaiC hexamers give rise to oscillations. KaiA binds to KaiC during the "day" and promotes phosphorylation, whereas at "night", KaiB binds to KaiC and sequesters KaiA thus leading to dephosphorylation. The horizontal rungs in all the states, correspond to the phosphotransfer reactions and the hydrolysis of ATP accompanying it, i.e. the \textit{red} arrows between $IA \longrightarrow IIB$, $IB \longrightarrow IIC$, \textit{purple} arrows between $IIIC \longrightarrow IIIA$, \textit{green} arrows between $IVD \longrightarrow IVA$ in Fig. \ref{MonomerCycle}. The ratio of the forward and backward rates is given by, $\gamma, \gamma_1, \gamma_2$ which are all less than 1, because of the fact that these describe reactions coupled to ATP hydrolysis which are inherently irreversible. In the model, $\alpha>1$ is responsible for differential affinity, $K_{d0} \equiv \frac{k_{Ab,0}}{k_{Af}}$ corresponds to \% ATP and $k_1$ helps in tuning ultrasensitivity. Free KaiA, $A_f$ provides non linearity to the system.}
    \label{MonomerCycleAndModel}
\end{figure}
\end{widetext}

The KaiC protein, complexed with KaiA, and KaiB proteins forms the core of the KaiABC oscillator system. The various possible states of the KaiC protein are described in Fig.~\ref{KaiCMonomer}. 
Our minimal model, described in Fig.~\ref{Model} and  inspired by Ref. \cite{Zhang2020} (with additional modifications to include features such as ultrasensitivity) can be viewed as a coarse-grained description of the various biochemical states accessed by the KaiABC protein system~\cite{Paijmans2017}. In the full KaiABC cycle, the KaiABC has two conformations, an active conformation (cyan background in Fig. \ref{MonomerCycleAndModel}) which can phosphorylate the Ser and Thr sites with KaiA as an assistant molecule and an inactive conformation (red background in Fig. \ref{MonomerCycleAndModel}) which sequesters KaiA with the help of KaiB and dephosphorylates the Thr and Ser sites. In our model, the $P_1$ and $P_3$ states correspond to the active conformation and $P_2$ to the inactive conformation.

The various biochemical states of the KaiABC protein are summarized in Fig.~\ref{KaiCMonomer} and Fig.~\ref{MonomerCycleAndModel}. Below, we briefly recap the various salient features of the KaiABC oscillatory cycle  and explain how they are taken into account in our minimal model.

\subsection{Differential binding of KaiA to KaiC drives the phosphorylation phase }
At the beginning of the cycle, most of the KaiC is in the active conformation in $CI_{DP}-CII^U_{DP}$ form ($IIIA$ in Fig. \ref{MonomerCycle}, $P_1(0)$ in Fig. \ref{Model}) and most of the KaiA is free. Depending on the phosphorylation level of active KaiC, it binds differently with KaiA. At low levels of phosphorylation ($IIIA, IIIB$) KaiC binds very strongly with KaiA. By constrast, the affinity of KaiA for KaiC is low when the KaiC is in a highly phosphorylated state ($IIIC, IIID$). This phenomena is termed as \textit{differential affinity} of KaiC for KaiA dimers \cite{vanZon7420}. Our model captures this effect through the parameter $\alpha$, where $\alpha>1$. Specifically, the rates of $P_1-P_3$ exchange are given by $k_{Af}A_f$ (where $A_f$ is the free KaiA concentration) from $P_1$ to $P_3$ and by $k_{Ab,0} \alpha^{\phi - \pi}$ in the reverse direction. As phosphorylation level increases with $\phi$, the term $\alpha^{\phi - \pi}$ ensures that proportion of $P_1$ (KaiA unbounded) states increases. The extent of differential affinity in our model can be tuned by varying the parameter $\alpha$. Differential affinity ensures that the unphosphorylated IIIA state is primed for KaiA binding at the start of the phosphorylation cycle. Indeed, KaiA binding to the $IIIA$ state transitions the system into the $IIA$ and $IA$ states. Subsequently, KaiA facilitates rapid exchange of nucleotides which lead to formation of more ATP bound states and pushes the system towards phosphorylation i.e. it leads to the formation of $CI_{TP}-^A CII_{TP}^S$, $CI_{TP}-^A CII_{TP}^T$ and $CI_{TP}-^A CII_{TP}^D$ states ($IB, IC$ and $ID$ states respectively in the schematic).

\subsection{Dependence of the kinetic rates on the ATP concentration}
The concentration of the energy rich molecule, ATP, is an important external condition for the cyanobacteria which affect the KaiABC oscillator. It has been observed that oscillations with almost the same time period are sustained till \%ATP in the system reaches 25\% below which oscillations vanish completely \cite{Phong2013}.  Here \%ATP $\equiv\frac{[ATP]}{[ATP] + [ADP]}$. In our model, the concentration of ATP controls the kinetics of the crucial $ATP-ADP$ nucleotide exchange reaction~\cite{Paijmans2017}.  Since in our minimal model the reaction corresponding to $III(A,B,C) \longrightarrow I(A,B,C)$ is coarse grained into, $P_1(i) \longrightarrow P_3(i)$ and since the second step in these reactions i.e. $II(A,B,C) \longrightarrow III(A,B,C)$ is dependent on \%ATP, the of  \%ATP in our model is set by the ratio of the rates connecting the $P_1$ to the $P_3$ states,
\begin{equation}
    \label{ATPdef}
    K_{d0} \equiv \frac{k_{Ab,0}}{k_{Af}}.
\end{equation}
Increasing $K_{d0}$ decreases the rate of transitions to the $P_3$ form and thus corresponds to lower \%ATP and vice versa.

\subsection{Dynamics of the dephosphorylation phase}

In the hexamer, the dephosphorylation phase starts even before total phosphorylation of each and every monomer. Specifically, once the number of serine sites phosphorylated becomes larger than the number of threonine sites which are occupied, the KaiA dissociates from the complex, the KaiC transform into an inactive conformation and the dephosphorylation phase kicks off. This transition corresponds to $ID\longrightarrow IID$ in the schematic Fig. \ref{MonomerCycle} and to the vertical rungs between $P_1$ and $P_2$ states colored \textit{magenta} in our model Fig. \ref{Model}.

The dephosphorylation phase ($IVD \longrightarrow IVA$) is relatively simple. It does not require KaiA as an assistant molecule for the reactions. When the proportion of doubly phosphorylated KaiC ($ID, IID$) is high, KaiB binding to the CI domain of KaiC is triggered, $IID \longrightarrow IIID$. In our model the KaiB binding to KaiC is taken into account implicitly during the transition from $P_1$ to $P_2$ states. KaiB bound KaiC,  $^BCI_{DP}-CII_{DP}^D$ ($IIID$) sequesters KaiA i.e. binds to KaiA and makes it unavailable for active use. This is taken into account through the parameter $\epsilon_{seq}$ in our model which reduces the free KaiA in the system by an amount $\epsilon_{seq} \sum P_2$. The dephosphorylation proceeds through the serine sites and then the threonine sites. Dephosphorylation reactions occur through phosphotransfer\cite{NISHIWAKI201218030}. This corresponds to the system moving through the $P_2$ states in our model. As the reactions reach the completely dephosphorylated state $^{AB} CI_{DP}-CII_{DP}^U$ ($IVB$), the KaiABC complex starts dissociating into KaiC, KaiB and release free KaiA into the system ($IVB \longrightarrow IVA$). The connection between $P_2(0)$ and $P_1(0)$ in our model taken this dissociation step. This prepares the system for the next cycle. 

\subsection{Ultrasensitive response of KaiC phosphorylation to KaiA concentration}

\begin{figure}[thb]
    \centering
    \includegraphics[width = 0.4\textwidth]{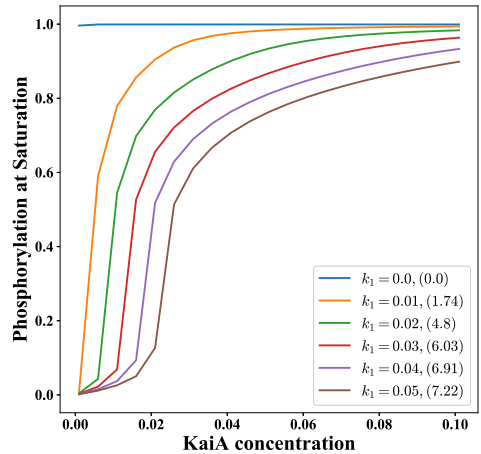}
    \caption{Ultrasensitive response in phosphorylation of KaiC wrt the total KaiA concentration for $K_{d0} = 10$ and $\alpha = 10$. The values in the bracket are the Hill Coefficients for the response curves (calculated using the method of relative amplification \cite{Legewie2005}). Values of other parameters are given in Table. \ref{ParameterValues}. These kinetics are in the absence of KaiB and $P_2$ states ($\omega = \omega_1 = 0$) i.e. they represent only the active form of KaiC in Fig. \ref{Model}. Thus there are no oscillations and the system always settles into a final steady state.}
    
    \label{Ultrasensitivity}
\end{figure}

It has been experimentally observed that in the absence of KaiB in the system, KaiC shows an ultrasensitive response in phosphorylation to KaiA concentration in the system i.e. the phosphorylation level of the KaiC hexamers change rapidly within a very narrow range of total KaiA concentration \cite{Phong2013, Hong2020}. This ultrasensitivity was speculated to be an important prerequisite for sustaining robust oscillations, particularly in conditions wherein the concentration of the energy rich molecule, ATP is low. Our model captures the ultrasensitive response observed in \cite{Hong2020} and described in Section.~\ref{modelsec}, through the introduction of the dephosphorylation rate $k_1$ (see Fig. \ref{Ultrasensitivity}). Indeed, a standard way to obtain ultrasensitive response is through the action of two antagonistic enzymes working at saturation \cite{Goldbeter6840,FerrelHa2014}. Under such conditions, the response of the system changes rapidly over a very narrow range of the enzyme concentration. In the KaiABC system the roles of the antagonistic enzymes are played by KaiA, which acts as a kinase phosphorylating KaiC and KaiC, which acts as its own phosphatase dephosphorylating itself \cite{Phong2013, NISHIWAKI201218030}.

The rate $k_1$ in our model captures this  dephosphorylation.  Tuning dephosphorylation rates by increasing $k_1$ leads to competition between phosphorylation in the $P_3$ states and dephoshporylation in the $P_1$ states. In the absence of KaiB, which corresponds to setting $\omega = \omega_1 = 0$ in our model, we consequently observe an ultrasensitive response of phosphorylation level of KaiC to changes in the KaiA concentration (Fig. \ref{Ultrasensitivity}).

\subsection{Dependence of the kinetic rates on the KaiA concentration}

As has been described above, the rates of transition between the $P_1$ and $P_3$ states in our minimal model depend on the concentration of free KaiA, $A_f$. The amount of free KaiA in turn depends on the concentrations of the $P_3$ and $P_2$ states since the KaiC complex is bound to KaiA in these states. Subsequently, $A_f=A_t - (\sum_{\phi} P_3(\phi) + \epsilon_{seq} \sum_{\phi} P_3(\phi)$. As the amount of $P_3$ and $P_2$ states increase, the free KaiA concentration decrease. This step gives rise to non-linearity in the system.

\renewcommand*{\thesection}{\Roman{section}}
\section{Role of differential affinity and ultrasensitivity: Insights from an analytical treatment of the Non-Linear Fokker Planck equations}
\renewcommand*{\thesection}{\arabic{section}}

Our minimal model described in Fig.~\ref{Model} and Sec.~\ref{modelsec} can be represented mathematically using a non-linear Fokker-Planck equation,$\frac{\partial \vec{P}}{\partial t} = \mathbf{W} \vec{P}$ where $\vec{P}$ is the probability vector of all the states ($P_1, P_2, P_3$) and $\mathbf{W} = \mathbf{W}(\vec{P})$ is the rate matrix dependent on the state of the system. The non linear Fokker-Planck equation is described in full detail in \ref{NonlinearFPEFull}.

If there were no nonlinearity in the Fokker-Planck equation, the Perron-Fobenius theorem would have ensured that the Fokker-Planck equation has a stable time independent steady-state solution. The oscillatory solutions of the rate-matrix decay with time as they have eigenvalues with a negative real part. Due to the non linearity in the Fokker-Planck equation in \eqref{ODEForAf}, time dependent oscillatory steady state solutions may be possible. 

In this work, we focus on how the solutions of the Fokker-Planck equation change as two specific parameters, namely, $\alpha$ controlling the differential affinity and $k_1$ controlling the ultransensitivity are varied. In particular, we analytically show how the system can be made to transition from a time independent steady state, where it cannot function as a biological clock, to a time dependent steady state, where it can function as a biological clock, as the differential affinity parameter $\alpha$ is tuned. For the case where the ultrasensitivity parameter $k_1$ is tuned, we take inspiration from our solution from tuning $\alpha$, and obtain an approximate solution. Our approximate analytical arguments provide insight into how ultrasensitivity also supports the functioning of the biological clock. 

Finally, as has been reported in many experimental and theoretical studies \cite{Phong2013, Tomita251, Paijmans2017}, oscillations are affected by the concentration of \%ATP in the system. In particular, it has been found that KaiABC system cannot sustain oscillations below a critical ATP concentration. In the next section, we will use our minimal model to show how stronger differential affinity and a better ultrasensitive switch can in fact sustain oscillations even at lower ATP concentrations \cite{Hong2020}. 

We begin our analytical treatment by first considering the case where $k_1=0$, i.e. in a model devoid of ultrasensitivity. In this case, a time independent solution for the non-linear Fokker-Planck equation can be obtained in the limit when $\epsilon_{seq} = 0$ and $\phi_0 = \pi$. $\epsilon_{seq} = 0$ corresponds to absence of KaiA sequestration by KaiB bound KaiC states. $\phi_0 = \pi$ means that the dephosphorylation phase starts only after all the KaiC have become doubly phosphorylated. Our analytical derivation is discussed in detail in ~\ref{k1equalto0calculation} and leads to the following solutions.
\begin{align}
\label{Casek1is0}
    &P_3(j) = b \ \forall j \in[0,N] \\
	&P_1(j) = \frac{1}{k_{Af}A_f} \left[ k_{Ab} \alpha^{\frac{j\pi}{N}} + (\delta_{0,j} - \delta_{j_0,j}) k_0 (1 - \gamma) \right] b \nonumber \\
	&\forall j \in[0,N] \label{Casek1iszero_P1}\\
	&P_2(2N - j) = \frac{k_0}{k_2}\left( \frac{1-\gamma}{1-\gamma_2} \right)b \nonumber \\
	&+ \gamma_2^j [ k_0(1-\gamma)\left( \frac{1}{\omega_1} - \frac{1}{k_2(1-\gamma_2)} \right)  \nonumber\\
	&+ \frac{K_D}{k_{Af} A_f}(k_0(1-\gamma) + k_{Ab}) ] \ \forall j \in[0,N]
\end{align}
where $ b = P_3(0)$  can be obtained by solving a quadratic equation as mentioned in \ref{TimeIndependentSteadyState}, $N = \frac{\pi}{\Delta \phi}, j = \frac{\phi}{\Delta \phi}, k_{Ab} = k_{Ab,0} \alpha^{-\pi}$. Even when $\phi_0 < \pi$, our solution gives a very good approximation if we set $P_1(j) \approx P_2(2N - j) \approx P_3(j) \approx 0 \ \forall j \in(j_0, N]$.

\begin{figure}[t]
    \def\big{\includegraphics[width=0.30\textwidth]{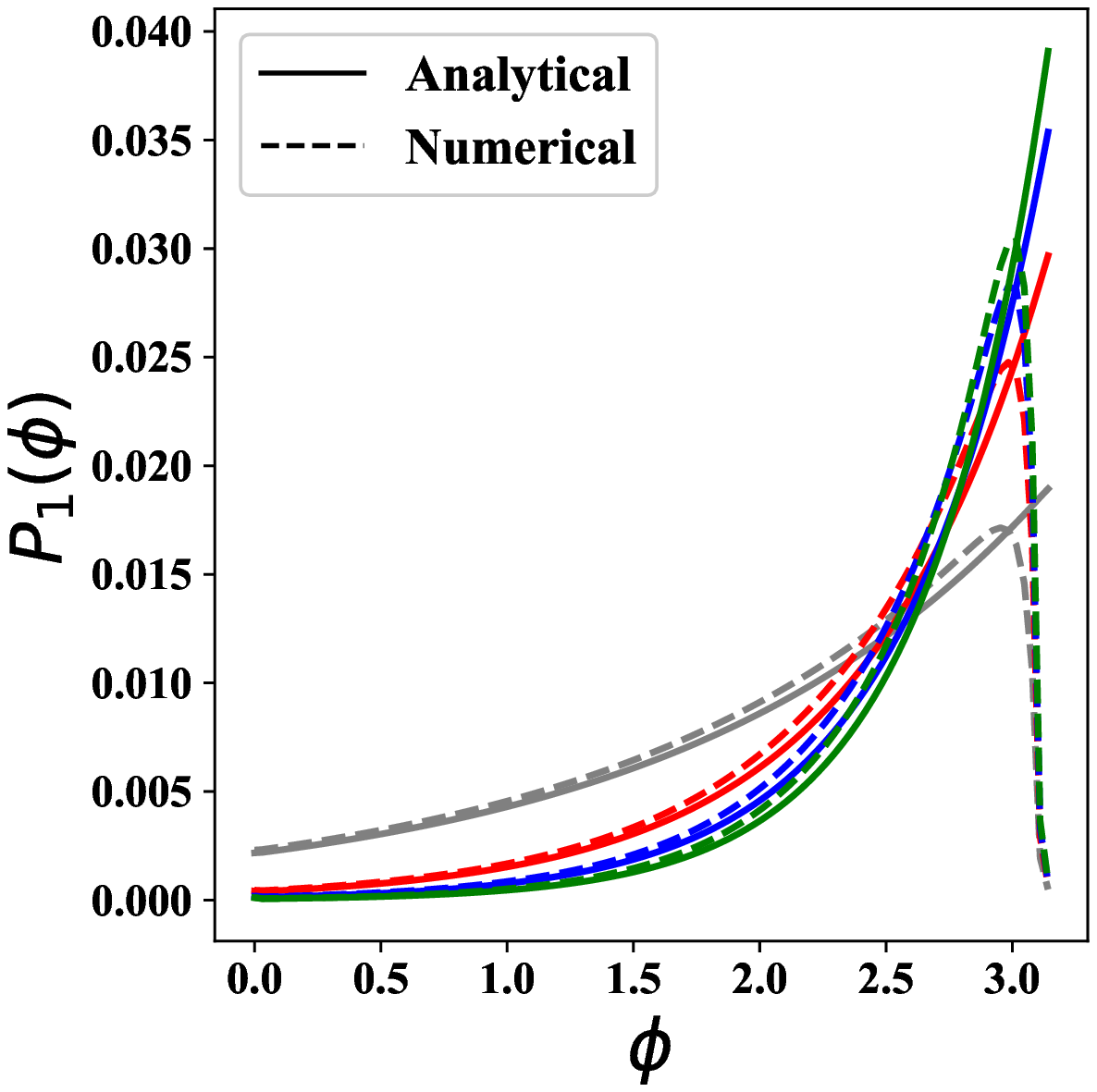}}
    \def\little{\includegraphics[width=0.15\textwidth]{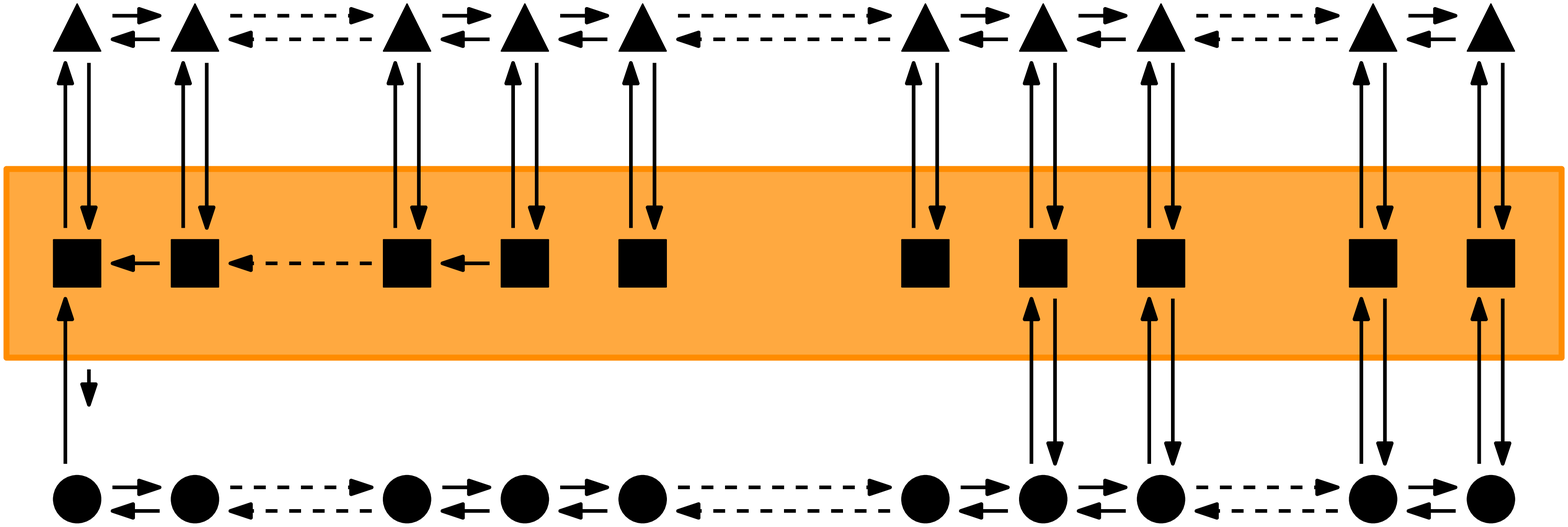}}
    \stackinset{l}{0.08\textwidth}{t}{0.1\textwidth}{\little}{\big}
    \caption{Comparison between numerical and analytical results for time-independent solution of $P_1$ states \eqref{Casek1iszero_P1} for different $\alpha$'s. The figure in inset is a representation of the Markov State network with the $P_1$ states highlighted. In the main figure, \textit{grey} corresponds to $\alpha = 2$, \textit{red} to $\alpha = 4$, \textit{blue} to $\alpha = 6$ and \textit{green} to $\alpha = 8$}
    \label{P1AnalyticalNumericalComparison}
\end{figure}

\begin{figure}[t]
    \centering
    \includegraphics[width = 0.5\textwidth]{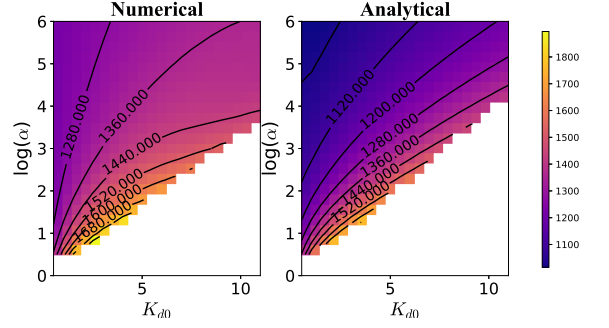}
    \caption{Time period of oscillations for various $\alpha$ and $K_{d0}$ i.e. at varying levels of differential affinity and \% ATP. $k_1 = 0$. Other parameters are given in Table \ref{ParameterValues_new}. Since, $k_1 = 0$, there is no effect of ultrasensitivity. The figure on the left represents time periods calculated by numerically simulating the FPE's. The figure on the right represents the time periods which were calculated from the imaginary part of the maximum positive eigenvalue of the instability matrix $W$, for small perturbations around the steady state probability distribution. As can be seen, the analytical solution provides us with a good approximation of the time period as well as the critical $\alpha$ at which oscillations take place for different $K_{d0}$. The contours in the figure are for the time period of oscillations.}
    \label{TimePeriodsAlphaKd0_paramset1}
\end{figure}

As $\alpha$ is increased, this time-independent state becomes unstable giving rise to a oscillatory state. As described in the \ref{LinearStabilityAnalysis}, a linear stability analysis can be performed around the steady state of the system, $\vec{P^s}$ to characterize this instability. The linear stability analysis has been detailed in \ref{OriginOfInstability}. This analysis correctly predicts the observed oscillatory behavior. Indeed, in Fig.~\ref{TimePeriodsAlphaKd0_paramset1} we show that the analytical estimate of the time period of oscillations provides a very good description of the actual observed oscillation periods. 

In the case of $k_1 \neq 0$ only an approximate solution for the time-independent steady state can be obtained. In order to obtain this approximate solution we take inspiration from the solution for the case when $k_1 = 0$ and assume $k_{Af}A_f P_1(\phi) = k_{Ab0} \alpha^{\phi - \pi} P_3(\phi)$ for $\phi \in [0, \phi_0]$ (along the $P_1 - P_3$ connections in Fig. \ref{Model}) and $P_1(\phi) \approx 0 \approx P_3(\phi)$ for $\phi>\phi_0$. This assumption is supported by numerical evidence. Under this assumption, we obtain,

\begin{align}
\label{Casek1isnot0}
    P_3(\phi) &= P_3(\phi') \frac{|B + A\alpha^{\phi'}|}{|B + A\alpha^{\phi}|} \ \forall \phi\in (0,\phi_0)\\
    P_1(\phi) &= \frac{K_{d0}}{A_f} \alpha^{\phi} P_3(\phi) \ \forall \phi\in (0,\phi_0)\\
    B &= -k_0 (1-\gamma)\Delta \phi, \ A = \frac{K_{d0}}{A_f}k_1 (1-\gamma_1)\Delta \phi
\end{align}
Here $P_3(0)$ can be obtained numerically and $\phi_0$ denotes the place where $P_1-P_2$ connections start in Fig. \ref{Model}. This is described in more detail in supplementary \ref{k1notzeroCalculation}. Fig. \ref{P3AnalyticalNumericalComparison} shows a comparison between the numerically obtained steady state with the one constructed using our approximate solution. We also provide approximate analytical arguments to show how a linear instability analysis can again be used to characterize the onset of oscillations as $k_1$ is tuned. The Gershgorin circle theorem provides us with a way to understand where we can find the eigenvalues of any matrix. As $k_1$ is tuned, the -ve off-diagonal elements of the rate matrix $\mathbf{W}$ increase in magnitude so do the radii of the Gershgorin discs (see Fig. \ref{Greshgorin}) because for any transition rate matrix, M, $\sum_{i} M_{ij} = 0$. In effect the Gershgorin discs have a finite area protruding into the positive half plane. With higher $k_1$ this area increases thus there is a higher chance of finding eigenvalues in the positive half-plane. These arguments are explained in more detail in \ref{OriginOfInstability}. 

\begin{figure}[ht]
    \def\big{\includegraphics[width=0.3\textwidth]{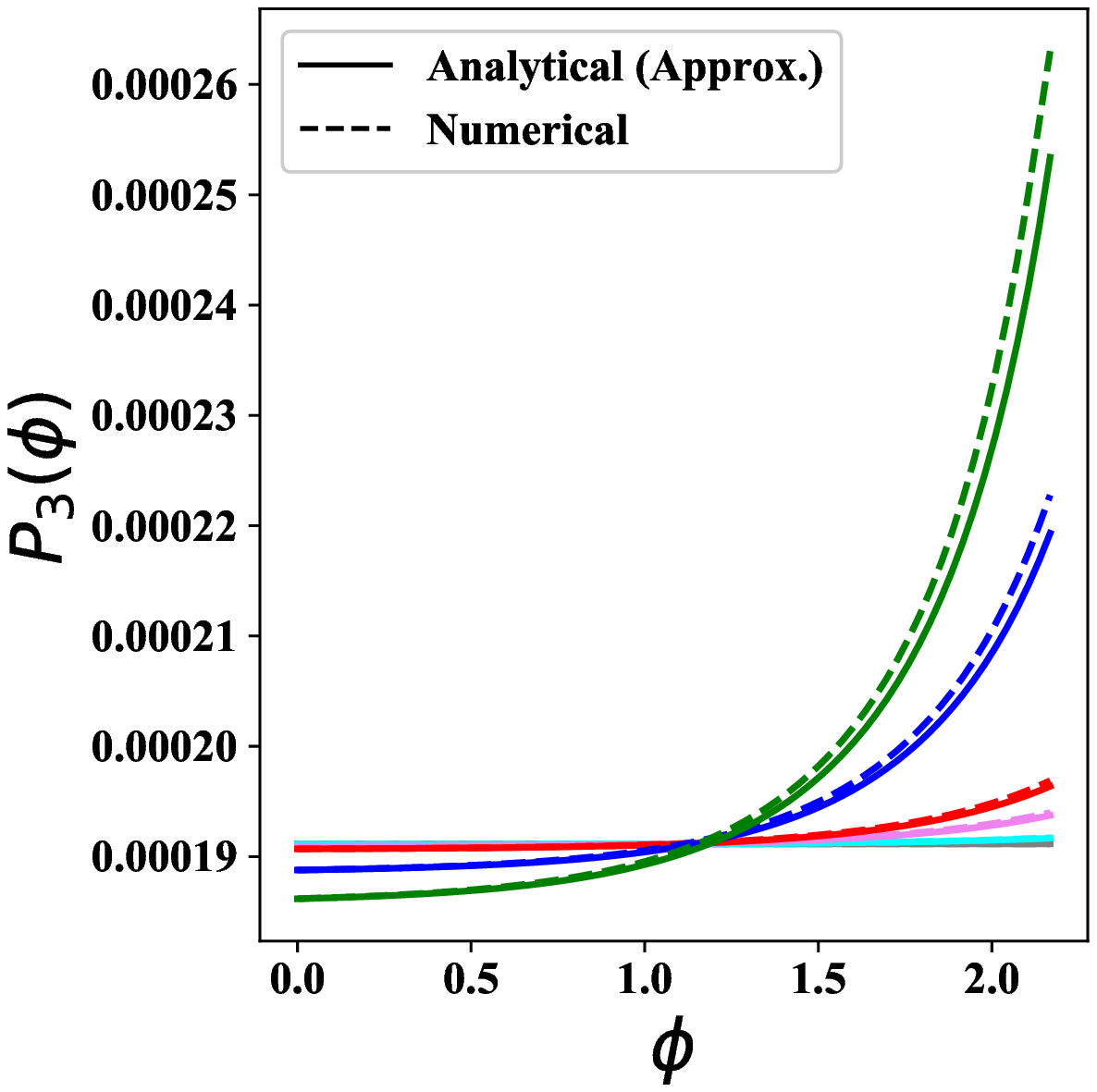}}
    \def\little{\includegraphics[width=0.15\textwidth]{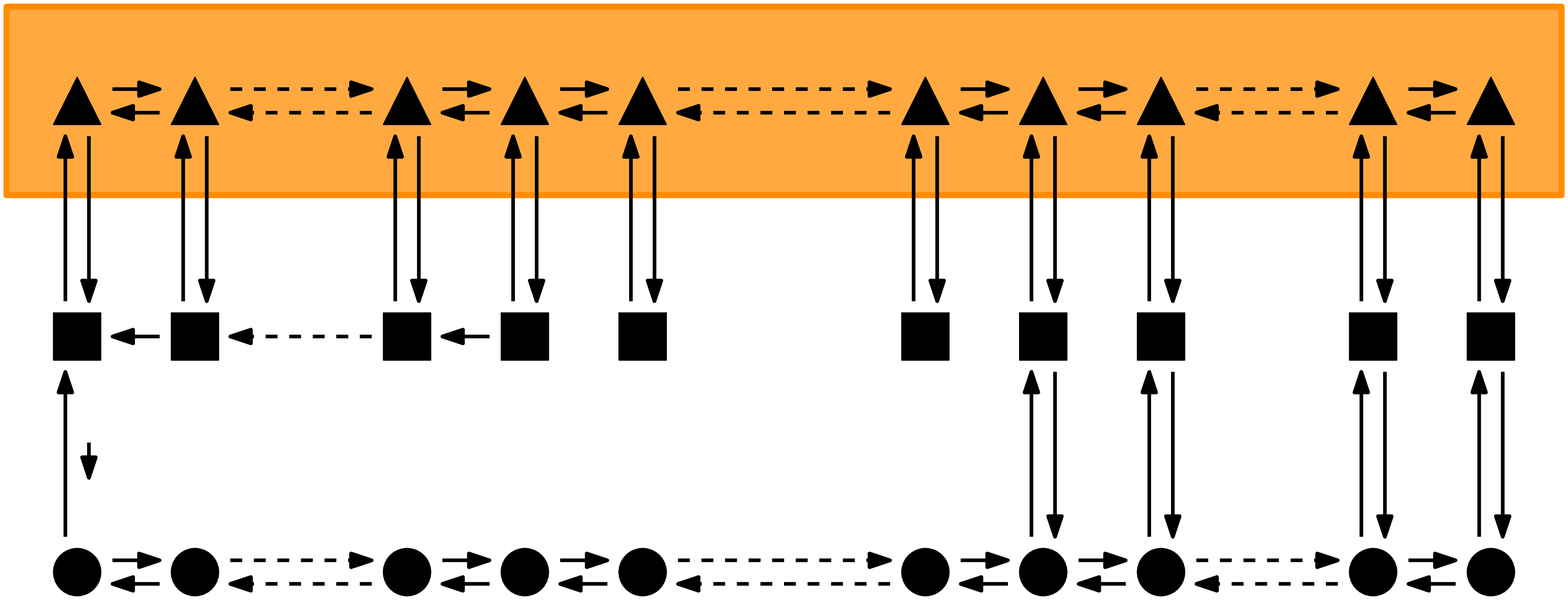}}
    \stackinset{l}{0.1\textwidth}{t}{0.1\textwidth}{\little}{\big}
    \caption{Comparison between numerical and approximate analytical results for time-independent solution of $P_3$ states for the case when $k_1\neq0$ \eqref{Casek1isnot0}. The figure in the inset represents the Markov State network with the $P_3$ states highlighted. In the main figure, \textit{grey} corresponds to $k_1 = 0$, \textit{cyan} to $k_1 = 10^{-4}$, \textit{violet} to $k_1 = 5\times10^{-4}$, \textit{red} to $k_1 = 10^{-3}$, \textit{blue} to $k_1 = 5\times10^{-3}$, \textit{green} to $k_1 = 10^{-2}$.}
    \label{P3AnalyticalNumericalComparison}
\end{figure}

\begin{figure}[thb]
    \begin{subfigure}[b]{0.45\textwidth}
    \includegraphics[width = \textwidth]{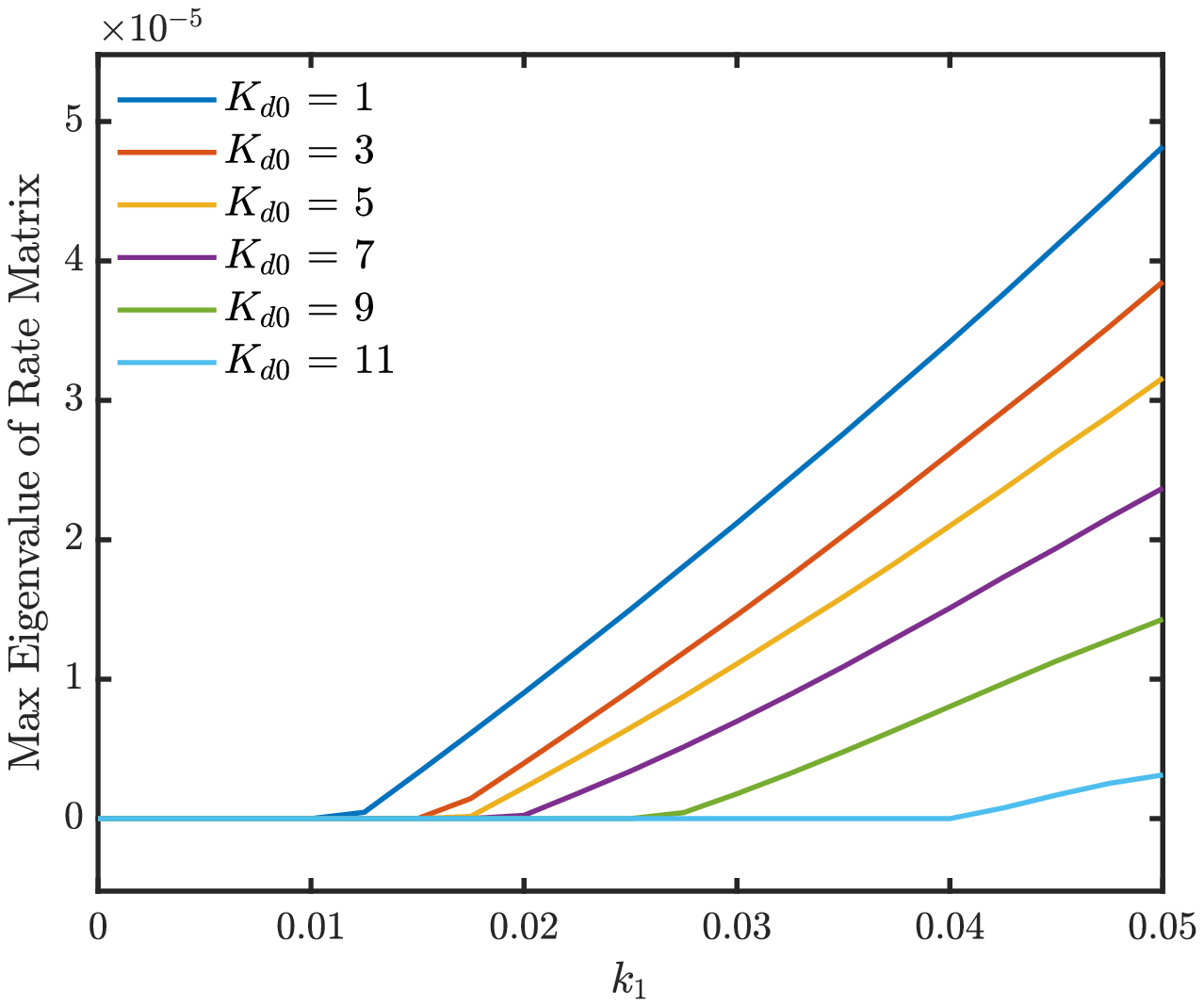}
    \end{subfigure}
    \begin{subfigure}[b]{0.2\textwidth}
    \centering
    \includegraphics[width = \textwidth]{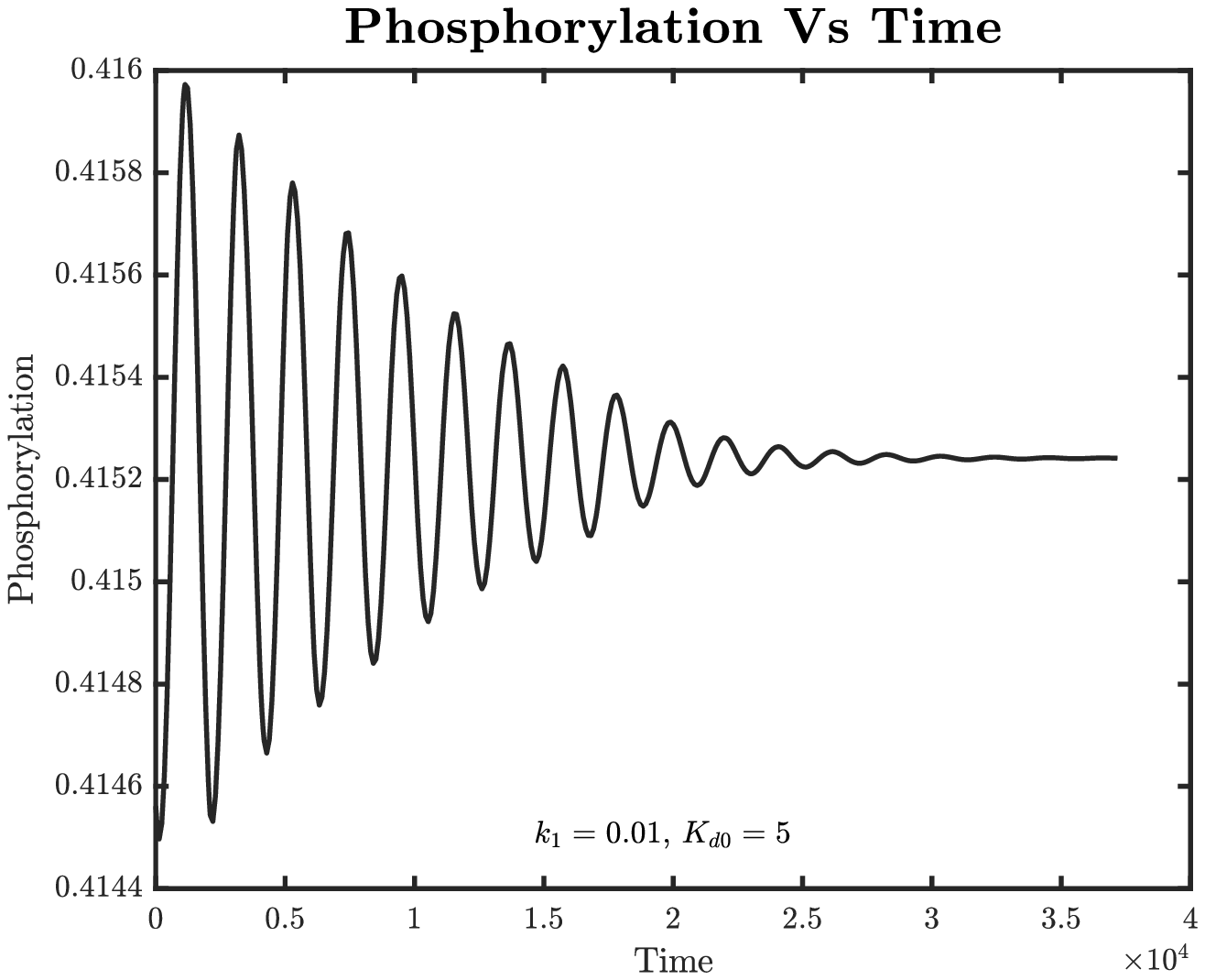}
    \caption{Damped Oscillations for $K_{d0} = 5$ and $k_1 = 0.01$}
    \end{subfigure}
    \begin{subfigure}[b]{0.2\textwidth}
    \centering
    \includegraphics[width = \textwidth]{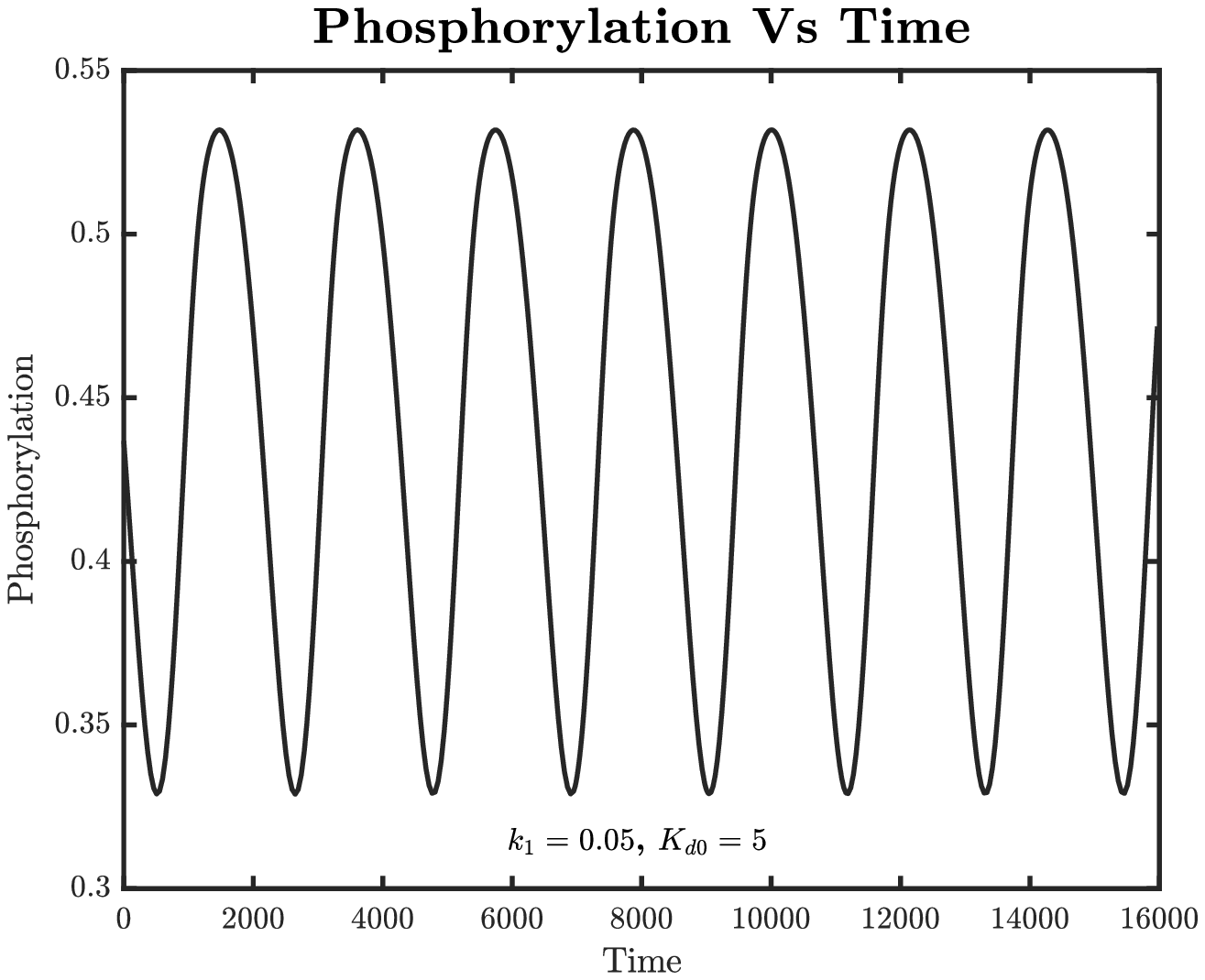}
    \caption{Oscillations for $K_{d0} = 5$ and $k_1 = 0.05$}
    \end{subfigure}
    \caption{Instability leading to oscillations when changing $k_1$. The y-axis denotes the maximum eigenvalue of the rate matrix W for the perturbations (refer to supplementary). The presence of a +ve eigenvalue denotes that the time-independent steady state is unstable. $\alpha = 10$ and the other parameter values are listed in Table \ref{ParameterValues}}
    \label{k1neq0Instability}
\end{figure}

In the next section, we build on these results and show how ultrasensitivity and differential affinity can support oscillations even at a lower ATP concentration. We also use insight from these analytical arguments to explain how the time period can be stably maintained in a variety of ATP concentrations, a phenomena known as affinity compensation. Finally, using our minimal model, we also comment on the thermodynamic costs associated with maintaining oscillations.

\renewcommand*{\thesection}{\Roman{section}}
\section{Discussion}
\renewcommand*{\thesection}{\arabic{section}}

\subsection{Increasing Differential Affinity leads to oscillations at low \%ATP}
\begin{figure}[tbb]
    \centering
    \includegraphics[width =0.32 \textwidth]{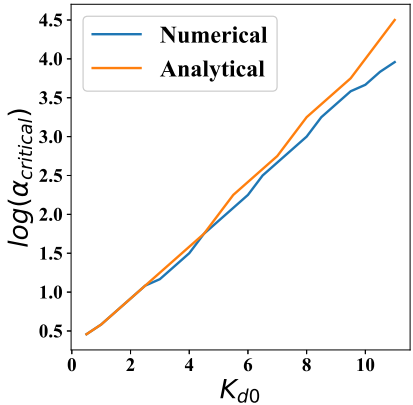}
    \caption{The value of $\alpha$ required for the onset of oscillations as a function of $K_{d0}$. Estimates have been obtained both from our theory and from numerical simulations. We set $k_1=0$ for these calculations.} .
    \label{CriticalAlpha}
    \end{figure}

It has been numerically shown previously in \cite{Zhang2020} that oscillations in a model system similar to ours can be obtained by increasing the value of $\alpha$ i.e. by improving the differential affinity. $\alpha$ controls the rate of reaction between $P_1$ and $P_3$ states in Fig. \ref{Model}. Our analytical results explain this numerical observation. Further, our analytical results at $k_1=0$ also help predict the required interplay between $\alpha$ and the ATP concentration in order for oscillations to be sustained. Specifically, we find that at $k_1 = 0$, a higher value of $\alpha$ is required for oscillations to take place at higher $K_{d0}$ (or a lower ATP concentration). In Fig. \ref{CriticalAlpha} we provide estimates of how the critical value of $\alpha$ changes as a function of the $K_{d0}$. Our analytical estimates agree very well with those obtained from the numerical calculations.

\subsection{Improving the ultrasensitive response leads to oscillations at lower \%ATP and fixed Differential Affinity}

\begin{figure}[t]
\centering
\includegraphics[width = 0.32\textwidth]{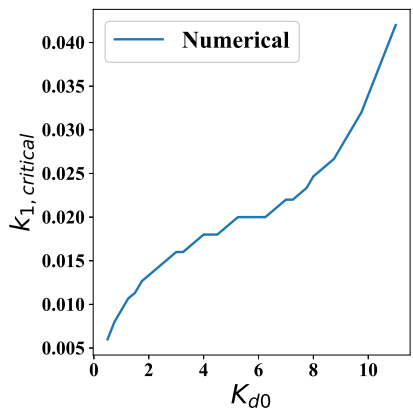}
\caption{The value of $k_1$ required for the onset of oscillations as a function of $K_{d0}$. Since the $k_1\neq 0$ is only approximately tractable analytically, we have only plotted estimates from numerical simulations.}
\label{Criticalk1}
\end{figure}

As mentioned in section \ref{modelsec}, it has been speculated that ultrasensitivity plays an important role in sustaining oscillations at low \%ATP conditions. Our minimal model captures this role played by ultransensitivity. Indeed, we find that at a higher value of $k_1$, corresponding to a sharper ultransensitive response (Fig. \ref{Ultrasensitivity}), oscillations can be sustained a larger $K_{d0}$ (or a smaller ATP concentration). We describe this tradeoff in Fig.\ref{k1neq0Instability} and Fig. \ref{Criticalk1}.

\begin{figure}[t]
    \centering
    \includegraphics[height = 0.32\textwidth, width = 0.35\textwidth]{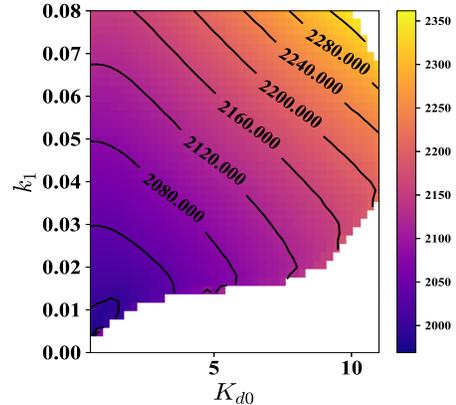}
    \caption{Time Period of oscillations for various $K_{d0}$ and $k_1$ i.e. at different levels of \% ATP and ultrasensitivity. The white region denotes the parameter space which does not support oscillations. This is also supported by the plot for the amplitude of oscillations, Fig. \ref{AmplitudeVsKd0k1}. In order to have oscillations at higher values of $K_{d0}$ the system requires a higher value of $k_1$. The contours in the figure are for the time period of oscillations.}
    \label{TimePeriodCompensation}
\end{figure}

\begin{figure}[t]
    \centering
    \includegraphics[height = 0.32\textwidth, width = 0.35\textwidth]{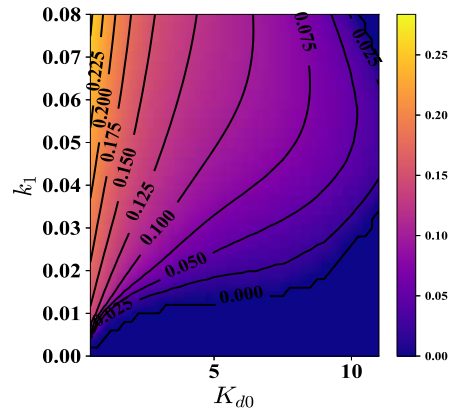}
    \caption{Amplitude of oscillations as a function of $K_{d0}$ and $k_1$ at $\alpha = 10$ and parameters given in \ref{ParameterValues}. The contours in the figure are for the amplitude of oscillations.}
    \label{AmplitudeVsKd0k1}
\end{figure}

Our analytical analysis also allows us to provide a phenomenological understanding of the role played by the ultransensitive switch. Ultrasensitivity offers coherence to the travelling wave-packet of phosphorylation at the start of every new cycle of oscillation. Phosphorylation is halted until a critical amount of KaiA is present in the system. Just before the beginning of every new phosphorylation cycle, most of the KaiA is sequestered by the $P_2$ states. Only after a certain amount of KaiA is freed from $P_2$ states, the phosphorylation reactions in the $P_3$ states can start again. This leads to a buildup of probability density near $P_2(2\pi)$ and $P_1(0)$ before the start of every cycle and provides coherence to the system and oscillations can be sustained.

\subsection{Metabolic compensation of Time period: Insights from the minimal Markov state model}
\begin{figure}[thb]
    \centering
    \includegraphics[width = 0.4\textwidth]{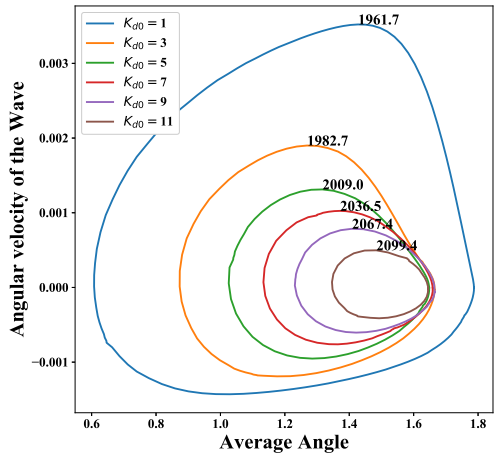} 
    \caption{Velocity of Phosphorylation wavepacket as a function of Average Angle for $k_1 = 0.05$, various $K_{d0}$'s and other parameters as given in Table. \ref{ParameterValues}. Here Average Angle, $\langle \phi \rangle = \sum_{\phi} \phi P(\phi)$ and Velocity, $v = \frac{d\langle\phi\rangle}{dt}$. The time period of oscillation for the different cycles is denoted along the curves.}
    \label{PhasePortrait}
\end{figure}

One of the most important feature of the KaiABC oscillator is that the time period of the oscillations are robust to changes in the \%ATP in the system, a phenomenon known as metabolic compensation. Our model shows a similar behaviour. Upon increasing $K_{d0}$, the time period increases, changing by 10\% for an increase from $K_{d0}=$1 to 11 (see Fig.~\ref{TimePeriodCompensation} and Fig.~\ref{PhasePortrait}). At $K_{d0}>11$, oscillations are not supported. This is analogous to losing oscillations when \%ATP is below 20\%ATP in the real system\cite{Phong2013, Paijmans2017}.

Our minimal model helps provide a simple phenomenological explanation of affinity compensation. In the regime where our model allows oscillations, the speed of the wave form as it traverses the top $P_1-P_3$ rungs in Fig.\ref{Model} from regions of lower $\phi$ to regions of higher $\phi$ can be shown to be $v = \frac{1}{3}\frac{(1-\gamma)k_0 - cK_{d0}(1-\gamma_1) k_1}{1 + cK_{d0}}$ through a first-passage time analysis (outlined in Sec.~\ref{FirstPassageTime}). Thus it is expected to decrease with $K_{d0}$. Simultaneously, $1/K_{d0}\equiv k_{Af}/k_{Ab,0}$ can be expected to control the relative occupancy of the $P_1-P_3$ states and the transitions in the $P3$ states promote probability flux towards regions of higher $\phi$. Thus, with increasing $1/K_{d0}$, the waveform can be expected to traverse more of the large $\phi$ states in the $P_3$ rung before transitioning to the $P1$ and then eventually to the $P_2$ states as it restarts the oscillation. Hence, at higher $1/K_{d0}$ or  higher \%ATP, the system traverses a larger orbit as described in the 'Angle-Angular Velocity' phase space (Fig. \ref{PhasePortrait}). This is analogous to shifting in the trough and crest in the phosphorylation oscillations observed in the KaiABC system \cite{Phong2013}. Together, these effects make the time period of oscillations relatively insensitive to \%ATP levels (Fig. \ref{PhasePortrait}). In this way, the KaiABC system can accomplish affinity compensation and maintain a relatively constant time period. 

\subsection{The thermodynamic costs of setting up oscillations}
\begin{figure}[ht]
    \centering
    \includegraphics[width = 0.35\textwidth]{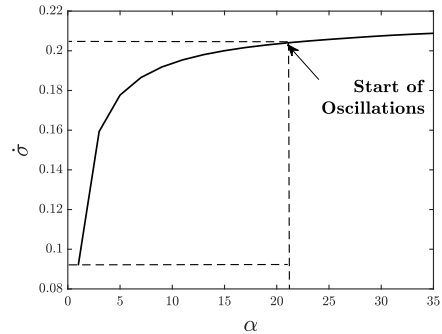}
    \vspace{0.01cm}
    \caption{Entropy Production Rate Vs $\alpha$ for $K_{d0} = 5, k_1 = 0$ and other parameters given in Table.\ref{ParameterValues_new}. Oscillations start at $\alpha = 21$. $\alpha = 1$ corresponds to absence of differential affinity. In order to have oscillations an additional 0.113 units of energy are required. This energy goes into building coherence among the KaiABC oscillator population \cite{Zhang2020}}
    \label{WorkRateAlphaKd05}
\end{figure}

\begin{figure}[ht]
    \centering
    \includegraphics[width = 0.35 \textwidth]{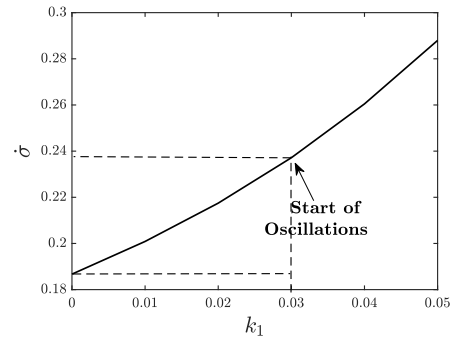}
    \caption{Entropy Production Rate Vs $k_1$ for $\alpha = 10, K_{d0} = 8$ and other parameters given in Table.\ref{ParameterValues}. Unlike the case with changing $\alpha$ in Fig. \ref{WorkRateAlphaKd05} where the entropy production plateaus very quickly with increasing $\alpha$, in this case, the entropy production increases almost linearly with increasing $k_1$. As expected, decreasing $K_{d0}$ and increasing $k_1$ lead to higher dissipation of energy. Oscillations start at $k_1 = 0.03$. $k_1 = 0$ corresponds to absence of ultrasensitivity in the system. An additional 0.052 units of energy are dissipated in order to have oscillations. This additional energy goes into improving the ultrasensitive response of the system eventually leading to coherence.}
    \label{RateOfWorkk1}
\end{figure}

Finally, the stochastic thermodynamics of our minimal model can be readily probed. The total steady state entropy production rate can be estimated using the probability fluxes along every edge of the model as \cite{Qian2007}
\begin{equation}
    \label{eq:entprod}
    \dot{\sigma} = \sum_{Edges} (J_+ - J_-)ln\frac{J_+}{J_-}
\end{equation}
We use Eq.~\ref{eq:entprod} to estimate the entropy production rate for various values of $\alpha$, $K_{d0}$ and $k_1$. These results are described in Fig. \ref{WorkRateAlphaKd05} and Fig. \ref{RateOfWorkk1}. Of particular note, our results show that $\dot{\sigma}$ varies continuously through the transition of the system from a stationary to an oscillatory phase. In the case where the ultrasensitivity parameter $k_1$ is tuned (Fig. \ref{RateOfWorkk1}), the entropy production rate, $\dot{\sigma}$ is almost a linearly increasing function of $k_1$. 
While the entropy production rate, $\dot{\sigma}$ does indeed increases as oscillations are setup in agreement with previous studies, ~\cite{Zhang2020}, and it does indeed improve the overall quality and coherence of oscillation \cite{Nyugen2018, Clara2020} an analysis focused on just the entropy production rate might miss the important and specific roles played by biophysical mechanisms such as the ultransensitivity and differential affinity in promoting and sustaining robust oscillations~\cite{Seara2021}.

\renewcommand*{\thesection}{\Roman{section}}
\section{Conclusion}
\renewcommand*{\thesection}{\arabic{section}}In conclusion, this work elucidates the role played by biophysical mechanisms such as ultrasensitivity and differential affinity in controlling the quality of circadian oscillations. Our minimal theoretical model also provides a route to explain how biochemical circuits can ensure oscillations with constant time periods, even under a range of experimental conditions. Finally, we show that the net rate of energy dissipation isn't a very effective order parameter to gauge the quality of oscillations, particularly in regimes where the ultrasensitivity is important. While our work relies on a very minimal abstraction of the KaiABC system. In future work we hope to adapt these ideas to more complex and complete models of circadian rhythm oscillators.

\bibliographystyle{unsrt}
\bibliography{Biblio_draft}

\begin{thebibliography}{10}

\bibitem{Mohawk2012}
Jennifer~A. Mohawk, Carla~B. Green, and Joseph~S. Takahashi.
\newblock Central and peripheral circadian clocks in mammals.
\newblock {\em Annual Review of Neuroscience}, 35(1):445--462, 2012.
\newblock PMID: 22483041.

\bibitem{Kondo2000}
Takao Kondo and Masahiro Ishiura.
\newblock The circadian clock of cyanobacteria.
\newblock {\em BioEssays}, 22(1):10--15, 2000.

\bibitem{BLAU2001287}
Justin Blau.
\newblock The drosophila circadian clock: what we know and what we don’t
  know.
\newblock {\em Seminars in Cell \& Developmental Biology}, 12(4):287--293,
  2001.

\bibitem{COLLINS2006348}
Ben Collins and Justin Blau.
\newblock Keeping time without a clock.
\newblock {\em Neuron}, 50(3):348--350, 2006.

\bibitem{Jeanne2005}
Jeanne~F. Duffy and Jr. Kenneth P.~Wright.
\newblock Entrainment of the human circadian system by light.
\newblock {\em Journal of Biological Rhythms}, 20(4):326--338, 2005.
\newblock PMID: 16077152.

\bibitem{Dubowy2017}
Christine Dubowy and Amita Sehgal.
\newblock {Circadian Rhythms and Sleep in Drosophila melanogaster}.
\newblock {\em Genetics}, 205(4):1373--1397, 04 2017.

\bibitem{MartinsE11415}
Bruno M.~C. Martins, Amy~K. Tooke, Philipp Thomas, and James C.~W. Locke.
\newblock Cell size control driven by the circadian clock and environment in
  cyanobacteria.
\newblock {\em Proceedings of the National Academy of Sciences},
  115(48):E11415--E11424, 2018.

\bibitem{Ouyang8660}
Yan Ouyang, Carol~R. Andersson, Takao Kondo, Susan~S. Golden, and Carl~Hirschie
  Johnson.
\newblock Resonating circadian clocks enhance fitness in cyanobacteria.
\newblock {\em Proceedings of the National Academy of Sciences},
  95(15):8660--8664, 1998.

\bibitem{Liaoe2022516118}
Yi~Liao and Michael~J. Rust.
\newblock The circadian clock ensures successful dna replication in
  cyanobacteria.
\newblock {\em Proceedings of the National Academy of Sciences}, 118(20), 2021.

\bibitem{Phong2013}
Connie Phong, Joseph~S. Markson, Crystal~M. Wilhoite, and Michael~J. Rust.
\newblock {Robust and tunable circadian rhythms from differentially sensitive
  catalytic domains}.
\newblock {\em Proceedings of the National Academy of Sciences of the United
  States of America}, 110(3):1124--1129, 2013.

\bibitem{Clodong2007}
Sébastien Clodong, Ulf Dühring, Luiza Kronk, Annegret Wilde, Ilka Axmann,
  Hanspeter Herzel, and Markus Kollmann.
\newblock Functioning and robustness of a bacterial circadian clock.
\newblock {\em Molecular Systems Biology}, 3(1):90, 2007.

\bibitem{AVELLO2021110495}
Paula Avello, Seth~J. Davis, and Jonathan~W. Pitchford.
\newblock Temperature robustness in arabidopsis circadian clock models is
  facilitated by repressive interactions, autoregulation, and three-node
  feedbacks.
\newblock {\em Journal of Theoretical Biology}, 509:110495, 2021.

\bibitem{DOVZHENOK20151830}
Andrey A. Dovzhenok, Mokryun Baek, Sookkyung Lim, and Christian I. Hong.
\newblock Mathematical modeling and validation of glucose compensation of the
  neurospora circadian clock.
\newblock {\em Biophysical Journal}, 108(7):1830--1839, 2015.

\bibitem{Paijmans2017}
Joris Paijmans, David~K. Lubensky, and Pieter~Rein ten Wolde.
\newblock A thermodynamically consistent model of the post-translational kai
  circadian clock.
\newblock {\em PLOS Computational Biology}, 13(3):1--43, 03 2017.

\bibitem{Rust220}
Michael~J. Rust, Susan~S. Golden, and Erin~K. O{\textquoteright}Shea.
\newblock Light-driven changes in energy metabolism directly entrain the
  cyanobacterial circadian oscillator.
\newblock {\em Science}, 331(6014):220--223, 2011.

\bibitem{Tomita251}
Jun Tomita, Masato Nakajima, Takao Kondo, and Hideo Iwasaki.
\newblock No transcription-translation feedback in circadian rhythm of kaic
  phosphorylation.
\newblock {\em Science}, 307(5707):251--254, 2005.

\bibitem{Hong2020}
Lu~Hong, Danylo~O Lavrentovich, Archana Chavan, Eugene Leypunskiy, and Eileen
  Li.
\newblock {Bayesian modeling reveals metabolite-dependent ultrasensitivity in
  the cyanobacterial circadian clock}.
\newblock {\em Molecular Systems Biology}, pages 1--23, 2020.

\bibitem{Hatakeyama2015}
Tetsuhiro~S. Hatakeyama and Kunihiko Kaneko.
\newblock {Reciprocity between robustness of period and plasticity of phase in
  biological clocks}.
\newblock {\em Physical Review Letters}, 115(21):1--5, 2015.

\bibitem{Nishiwaki13927}
Taeko Nishiwaki, Yoshinori Satomi, Masato Nakajima, Cheolju Lee, Reiko
  Kiyohara, Hakuto Kageyama, Yohko Kitayama, Mioko Temamoto, Akihiro Yamaguchi,
  Atsushi Hijikata, Mitiko Go, Hideo Iwasaki, Toshifumi Takao, and Takao Kondo.
\newblock Role of kaic phosphorylation in the circadian clock system of
  synechococcus elongatus pcc 7942.
\newblock {\em Proceedings of the National Academy of Sciences},
  101(38):13927--13932, 2004.

\bibitem{Rust16760}
Michael~J. Rust.
\newblock Orderly wheels of the cyanobacterial clock.
\newblock {\em Proceedings of the National Academy of Sciences},
  109(42):16760--16761, 2012.

\bibitem{Zhang2020}
Dongliang Zhang, Yuansheng Cao, Qi~Ouyang, and Yuhai Tu.
\newblock {The energy cost and optimal design for synchronization of coupled
  molecular oscillators}.
\newblock {\em Nature Physics}, 16(1):95--100, 2020.

\bibitem{NISHIWAKI201218030}
Taeko Nishiwaki and Takao Kondo.
\newblock Circadian autodephosphorylation of cyanobacterial clock protein kaic
  occurs via formation of atp as intermediate*.
\newblock {\em Journal of Biological Chemistry}, 287(22):18030--18035, 2012.

\bibitem{vanZon7420}
Jeroen~S. van Zon, David~K. Lubensky, Pim R.~H. Altena, and Pieter~Rein ten
  Wolde.
\newblock An allosteric model of circadian kaic phosphorylation.
\newblock {\em Proceedings of the National Academy of Sciences},
  104(18):7420--7425, 2007.

\bibitem{Goldbeter6840}
A~Goldbeter and D~E Koshland.
\newblock An amplified sensitivity arising from covalent modification in
  biological systems.
\newblock {\em Proceedings of the National Academy of Sciences},
  78(11):6840--6844, 1981.

\bibitem{FerrelHa2014}
James E~Ferrell Jr and Sang~Hoon Ha.
\newblock {Ultrasensitivity part I : Michaelian responses and zero-order
  ultrasensitivity}.
\newblock {\em Trends in Biochemical Sciences}, 39(10):496--503, 2014.

\bibitem{Legewie2005}
Stefan Legewie, Nils Bl{\"{u}}thgen, and Hanspeter Herzel.
\newblock {Quantitative analysis of ultrasensitive responses}.
\newblock {\em FEBS Journal}, 272(16):4071--4079, 2005.

\bibitem{Qian2007}
Hong Qian.
\newblock Phosphorylation energy hypothesis: Open chemical systems and their
  biological functions.
\newblock {\em Annual Review of Physical Chemistry}, 58(1):113--142, 2007.
\newblock PMID: 17059360.

\bibitem{Nyugen2018}
Basile Nguyen, Udo Seifert, and Andre~C. Barato.
\newblock Phase transition in thermodynamically consistent biochemical
  oscillators.
\newblock {\em The Journal of Chemical Physics}, 149(4):045101, 2018.

\bibitem{Clara2020}
Clara del Junco and Suriyanarayanan Vaikuntanathan.
\newblock High chemical affinity increases the robustness of biochemical
  oscillations.
\newblock {\em Phys. Rev. E}, 101:012410, Jan 2020.

\bibitem{Seara2021}
Daniel~S. Seara, Benjamin~B. Machta, and Michael~P. Murrell.
\newblock {Irreversibility in dynamical phases and transitions}.
\newblock {\em Nature Communications}, 12(1):1--9, 2021.

\end{thebibliography}

\renewcommand\thesection{\arabic{section}}
\clearpage
\onecolumngrid

\section*{\LARGE Supplementary Information}

\setcounter{figure}{0}
\setcounter{table}{0}
\setcounter{equation}{0}
\setcounter{page}{1}
\setcounter{section}{0}

\renewcommand{\theequation}{S\arabic{equation}} 
\renewcommand{\thepage}{S\arabic{page}} 
\renewcommand{\thesection}{S\arabic{section}}  
\renewcommand{\thetable}{S\arabic{table}}  
\renewcommand{\thefigure}{S\arabic{figure}}

\numberwithin{equation}{section}

\section{Model Details}
\label{NonlinearFPEFull}
\begin{figure}[H]
\centering
\includegraphics[width = 0.6\textwidth]{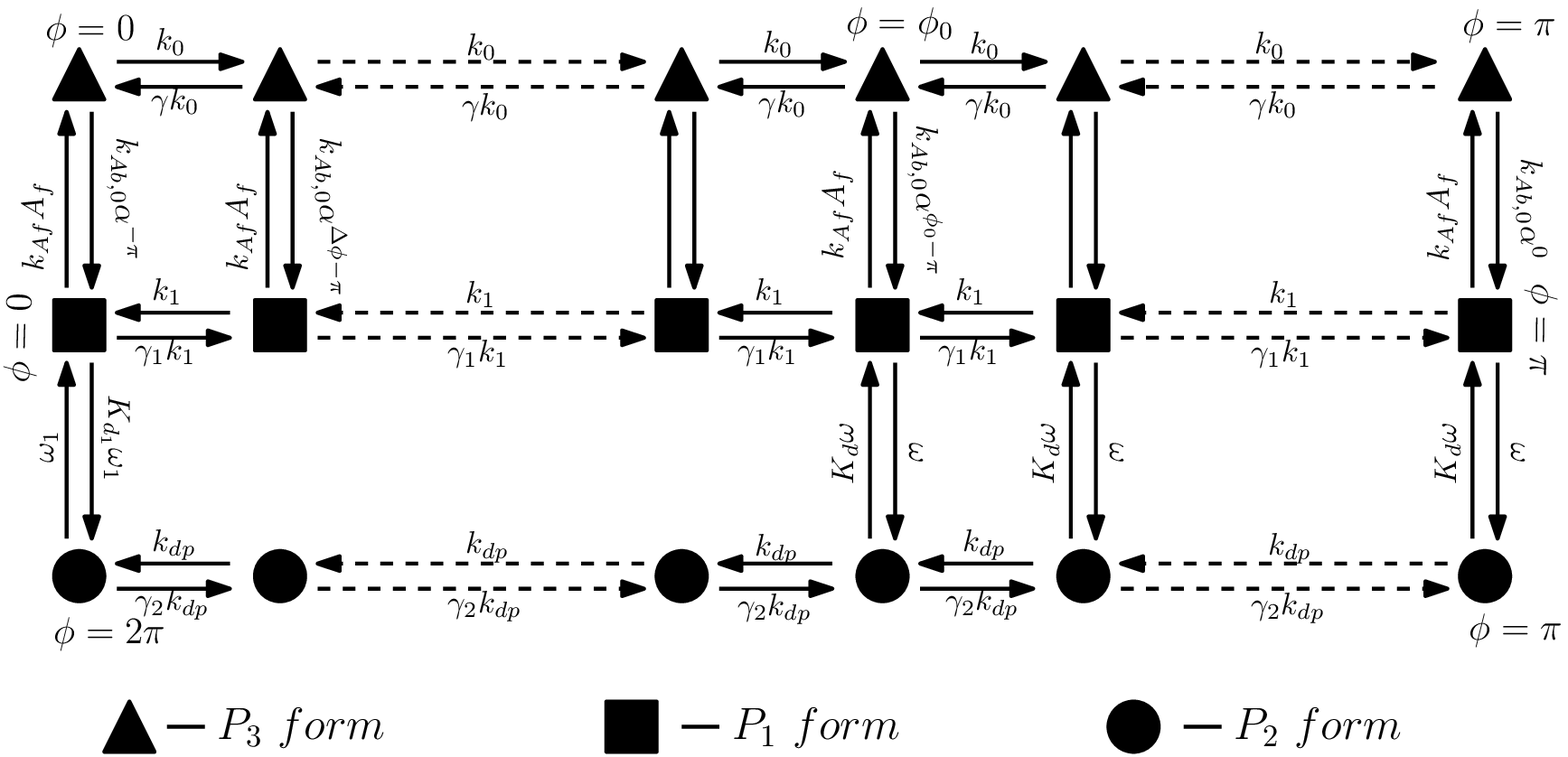}
\caption{Each oscillator can exist in one of the three forms	$P_1, P_2, P_3$. The $P_1$ form corresponds to normal KaiC during daytime and requires KaiA for moving forward in phase. The $P_3$ form corresponds to the form of KaiC to which KaiA is attached as an assistant molecule, $P_1 + KaiA \longrightarrow P_3$. The $P_2$ form corresponds to KaiC in it's dephosphorylation phase. The nucleotide bound states, KaiB binding and KaiA sequestration are implicitly assumed in the model. $A_f$ denotes the free KaiA concentration and $A_t$ stands for the total KaiA concentration. Phosphorylation corresponds to phase $\phi$, it increases linearly from 0 to 1 as $\phi$ varies from 0 to $\pi$ and decreases linearly from 1 to 0 as $\phi$ goes from $\pi$ to $2\pi$. Connections between the $P_1$ states denote the spontaneous dephosphorylation of KaiC subunits in the absence of KaiA. Multiple connections between $P_1$ and $P_2$ towards the highly phosphorylated states allows the system to move to the dephosphorylation phase even before all the KaiC is completely phosphorylated. The $P_2$ states sequester KaiA which corresponds to the fact that during the dephosphorylation phase, KaiB bound KaiC sequesters KaiA and makes it inactive. The master equation for this system is given below.}
\end{figure}

\begin{align}
\frac{\partial P_1(\phi)}{\partial t} &= \delta(\phi \neq \pi)k_1(P_1(\phi+\Delta\phi) - \gamma_1P_1(\phi)) + \delta(\phi \neq 0)k_1(\gamma_1P_1(\phi - \Delta \phi) - P_1(\phi)) \nonumber\\
&+\delta(\phi)(\omega_1 P_2(2\pi - \phi) - K_{d1} \omega_1 P_1(\phi)) \nonumber\\
&+ H(\phi - \phi_0)(K_d \omega P_2(2\pi - \phi) - \omega P_1(\phi)) -k_{Af} A_f P_1(\phi) + k_{Ab,0} \alpha^{\phi - \pi} P_3(\phi)  \label{ODEForP1} \\
\frac{\partial P_2(\phi)}{\partial t} &= \delta(\phi \neq 2\pi)k_{dp}(\gamma_2P_2(\phi+\Delta\phi) - P_2(\phi)) + \delta(\phi \neq \pi)k_{dp}(P_2(\phi - \Delta \phi) - \gamma_2P_2(\phi)) \nonumber\\
&-\delta(\phi-2\pi)(\omega_1 P_2(\phi) - K_{d1} \omega P_1(\phi - 2\pi))
+H(2\pi - \phi - \phi_0)(\omega P_1(2\pi - \phi) - \omega P_2(\phi)) \\
\frac{\partial P_3(\phi)}{\partial t} &= \delta(\phi \neq \pi)k_{0}(\gamma P_3(\phi+\Delta\phi) - P_3(\phi)) + \delta(\phi \neq 0)k_{0}(P_3(\phi - \Delta \phi) - \gamma P_3(\phi)) \nonumber\\
&+k_{Af} A_f P_1(\phi) - k_{Ab,0} \alpha^{\phi - \pi} P_3(\phi) \\
A_f &= A_t - \sum_0^{\pi}P_3(\phi) - \epsilon_{seq}\sum_{\pi}^{2\pi}P_2(\phi), \ \epsilon_{seq}<A_t \\
\frac{d A_f}{dt} &= \sum_0^{\pi} (k_{Ab,0} \alpha^{\phi - \pi} P_3(\phi)-k_{Af} A_f P_1(\phi)) \nonumber\\
&+ \epsilon_{seq} \sum_{\pi}^{2\pi}( H(2\pi - \phi - \phi_0) \omega P_2(\phi) - K_d \omega P_1(2\pi - \phi)) + \epsilon(\omega_1 P_2(2\pi) - K_{d1} \omega_1 P_1(0)) 5\label{ODEForAf} \\
H(\phi) &= 1 \ for \ \phi \geq 0 \ and \ 0 \ for \ \phi<0
\end{align}

For the sake of convenience, we relabel the $P_2$ states such that $P_2(2N - j) \Longleftrightarrow P_2(j)$. We also work with the discrete case so we relabel the $\phi$ using j, where $j = \frac{\phi}{\Delta \phi}$. Relabelling does not change the dynamics. The Fokker-Planck equations in this case are given by,
\begin{align}
	\frac{\partial P_1(j)}{\partial t} &= \delta(j \neq N)k_1(P_1(j+1) - \gamma_1P_1(j)) + \delta(j \neq 0)k_1(\gamma_1P_1(j - 1) - P_1(j)) \nonumber\\
	&+\delta(j=0)(\omega_1 P_2(j) - K_{d1} \omega_1 P_1(j)) \nonumber\\
	&+ H(j - j_0)(K_d \omega P_2(j) - \omega P_1(j)) -k_{Af} A_f P_1(j) + k_{Ab} \alpha^{j} P_3(j)  \\
	\frac{\partial P_2(j)}{\partial t} &= \delta(j \neq 0)k_{dp}(\gamma_2P_2(j-1) - P_2(j)) + \delta(j \neq N)k_{dp}(P_2(j + 1) - \gamma_2P_2(j)) \nonumber\\
	&-\delta(j=0)(\omega_1 P_2(j) - K_{d1} \omega P_1(j)) + H(j - j_0)(\omega P_1(j) - \omega P_2(j)) \\
	\frac{\partial P_3(j)}{\partial t} &= \delta(j \neq N)k_{0}(\gamma P_3(j+1) - P_3(j)) + \delta(j \neq 0)k_{0}(P_3(j - 1) - \gamma P_3(j)) \nonumber\\
	&+k_{Af} A_f P_1(j) - k_{Ab} \alpha^{j} P_3(j) \\
	A_f &= A_t - \sum_0^{N}P_3(j) - \epsilon_{seq}\sum_{0}^{N}P_2(j)
\end{align}

\section{Time Independent Steady State}
\label{TimeIndependentSteadyState}
\subsection{Case I : $k_1=0$}
\label{k1equalto0calculation}
We set, $k_1 = 0, j_0 = N, K_{d1} = K_d = K_D$, we also make the following changes in notation $k_{Ab,0}\alpha^{-\pi} = k_{Ab}, k_{dp} = k_2, \alpha->\alpha^{\frac{\pi}{N}}$. Using this simplification, we can solve for the steady state solution of the system and then use linear stability analysis around the steady state to see how oscillations are set up.

\begin{figure}[H]
	\centering
	\includegraphics[width = 0.8\textwidth]{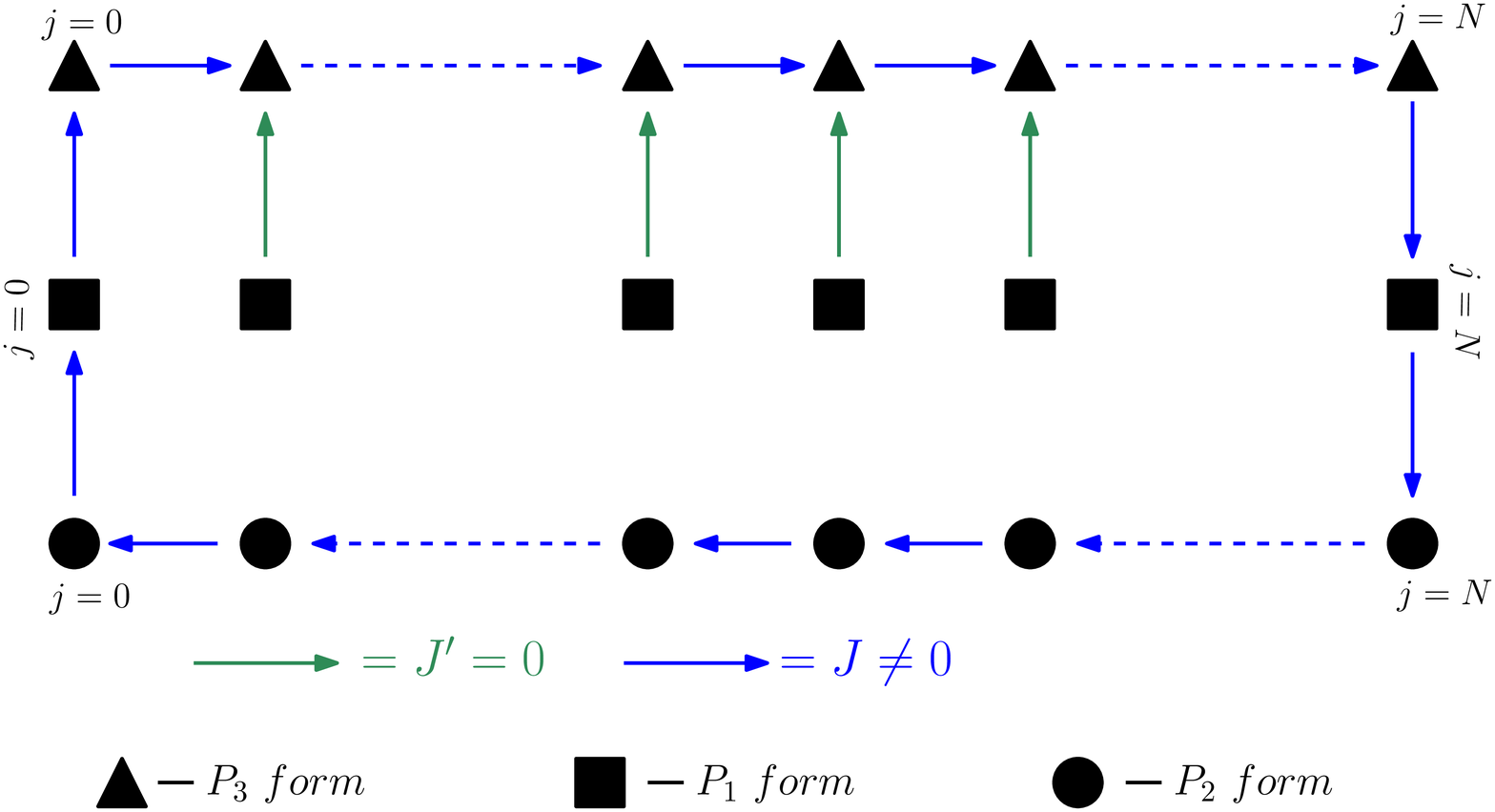}
	\caption{There is just one flux, J in the entire system. Once we solve for this unique flux J, we can obtain the expressions for the probability distribution of the different states in the system.}
	\label{SingleFlux}
\end{figure}

\begin{align}
	A_f &= A_t - \epsilon \sum_{j=0}^N P_3(j), \ P_3(0) = b, \ P_1(0) = a
\end{align}
where a and b are the labels for $P_1(0)$ and $P_3(0)$ respectively. For the $P_1 - P_3$ connection at $j=0$ and $j=N$ we have,
\begin{align}
	J &= k_{Af}A_fP_1(0) - k_{Ab}\alpha^0 P_3(0) = k_{Af}A_fa - k_{Ab}b \label{P1P3_egde0} \\
	J &= k_{Ab} \alpha^N P_3(N) - k_{Af} A_f P_1(N) \label{P1P3_edgeN}
\end{align}
For other $P_3-P_3$ connections, we have,
\begin{align}
	J &= k_0P_3(j-1) - \gamma k_0P_3(j) \\
	\implies P_3(j) &= \frac{1}{\gamma k_0}\left[k_0 P_3(j-1) - J \right] \\
	\therefore P_3(1) &= \frac{1}{\gamma k_0}\left[ k_0b - J \right] ,\ P_3(j) = \frac{1}{\gamma k_0} \left[ \frac{k_0}{\gamma^{j-1}}b - J\left( 1 + \frac{1}{\gamma} + ... + \frac{1}{\gamma^{j-1}} \right) \right] \\
	\implies P_3(j) &= \frac{1}{\gamma k_0} \left[ \frac{k_0}{\gamma^{j-1}}b - J\left( \frac{1 - \frac{1}{\gamma^j}}{1 - \frac{1}{\gamma}} \right) \right] \label{ExpressionForP3}
\end{align}
If we look at $P_1 - P_3$ connections in the bulk, we have,
\begin{align}
	J' &= 0 \implies k_{Af}A_fP_1(j) = k_{Ab}\alpha^j P_3(j) \ \forall \ j \neq 0, N \label{P1P3_bulk}
\end{align}

When we look at the $P_1 - P_2$ connection at $j=0$, we have,
\begin{align}
	J &= \omega_1 P_2(0) - K_D \omega_1 P_1(0) \implies P_2(0) = \frac{1}{\omega_1}\left[ J + K_{d1} \omega_1 a \right] = c 
\end{align}
where c is the label for $P_2(0)$. Similarly for other $P_2-P_2$ connections we have the following,
\begin{align}	
	J &= k_2P_2(j) - \gamma_2 k_2P_2(j-1) \\
	\implies P_2(1) &= \frac{1}{k_2} \left[ J + \gamma_2 k_2 c \right] ,\ P_2(j) = \frac{1}{k_2}\left[ \gamma_2^j k_2 c + J(1+\gamma_2 +...+\gamma_2^{j-1}) \right] \\
	\implies P_2(j) &= \frac{1}{k_2}\left[ \gamma_2^j k_2 c + J\frac{1 - \gamma_2^j}{1 - \gamma_2} \right]  \label{ExpressionForP2} \\
\end{align}
Substituting the expressions of $P_1(N)$ \eqref{P1P3_edgeN}, $P_2(N)$ \eqref{ExpressionForP2} and $P_3(N)$ \eqref{ExpressionForP3} into \eqref{AtEdgeN}, we get,
\begin{align}
	&At \ j=N, \ k_{Ab}\alpha^N P_3(N) - k_{Af}A_f P_1(N) = J = \omega P_1(N) - K_D \omega P_2(N) \label{AtEdgeN} \\
	\implies \frac{k_{Ab} \alpha^N}{\gamma^N} &- k_{Af}A_fK_d K_{d1} \gamma_2^N a = \left[ 1 + \frac{k_{Af}A_f}{\omega_1} + \frac{k_{Af}A_f K_d}{k_2} \frac{1- \gamma_2^N}{1 - \gamma_2} + \frac{k_{Ab} \alpha^N}{\gamma k_0} \frac{1 - \frac{1}{\gamma^N}}{1 - \frac{1}{\gamma}} + \frac{k_{Af} A_f K_d \gamma_2^N}{\omega} \right]J \label{main:eqn} \\
	J &= k_{Af}A_fa - k_{Ab}b
\end{align}
Since N is large and $\gamma < 1$, the LHS and RHS \eqref{main:eqn} are dominated by the terms having $\frac{1}{\gamma^N}$. Thus only the $1^{st}$ term in the LHS and the $3^{rd}$ term in the RHS of \eqref{main:eqn} contribute, other terms can be ignored. This leads to an expression for J.
\begin{align}
	\implies J &= k_0 (1-\gamma)b
\end{align}
Substituting this expression for J in \eqref{ExpressionForP3}, \eqref{ExpressionForP2}, \eqref{P1P3_egde0}, \eqref{P1P3_edgeN}, \eqref{AtEdgeN}, we get,
\begin{align}
	\therefore P_3(j) &= b \forall j \label{P3expression_k1_0}\\
	P_1(j) &= \frac{1}{k_{Af}A_f} \left[ k_{Ab} \alpha^j + (\delta_{0,j} - \delta_{N,j}) k_0 (1 - \gamma) \right] b \label{P1expression_k1_0}\\
	P_2(j) &= \frac{k_0}{k_2}\left( \frac{1-\gamma}{1-\gamma_2} \right)b + \gamma_2^j \left[ k_0(1-\gamma)\left( \frac{1}{\omega_1} - \frac{1}{k_2(1-\gamma_2)} \right)  + \frac{K_{d1} \omega}{k_{Af} A_f \omega_1}(k_0(1-\gamma) + k_{Ab}) \right] b \label{P2expression_k1_0}
\end{align}

Now using, $\sum_{j = 0}^{N} (P_1(j) + P_2(j) + P_3(j)) = 1$,
\begin{multline}
	\implies \frac{1}{b} = \frac{1}{k_{Af}A_f} \left[ k_{Ab}\frac{\alpha^{N+1} - 1}{\alpha - 1} + \frac{K_{d1} \omega (k_0 (1-\gamma) + k_{Ab})}{(1 - \gamma_2) \omega_1} \right] \\  
	+  (N+1)\left( 1 + \frac{k_0(1-\gamma)}{k_2 (1-\gamma_2)} \right) + \frac{1}{1 - \gamma_2}\left( k_0 \left(\frac{1- \gamma}{\omega_1}\right) -\frac{k_0(1-\gamma)}{k_2(1-\gamma_2)} \right)
\end{multline}

$A_f = A_t - \epsilon \sum_{j=0}^N P_3(j) - \epsilon_{seq} \sum_{j=0}^N P_3(j) = A_t + f - gb$, where f and g are constants. Setting $\epsilon_{seq} = 0$, we get, $A_f = A_t - (N+1)b$ and we need to solve a quadratic equation to find the probabilities which determine the steady state.

\begin{figure}[thb]
\begin{subfigure}[b]{0.37\textwidth}
\centering
\def\big{\includegraphics[width=\textwidth]{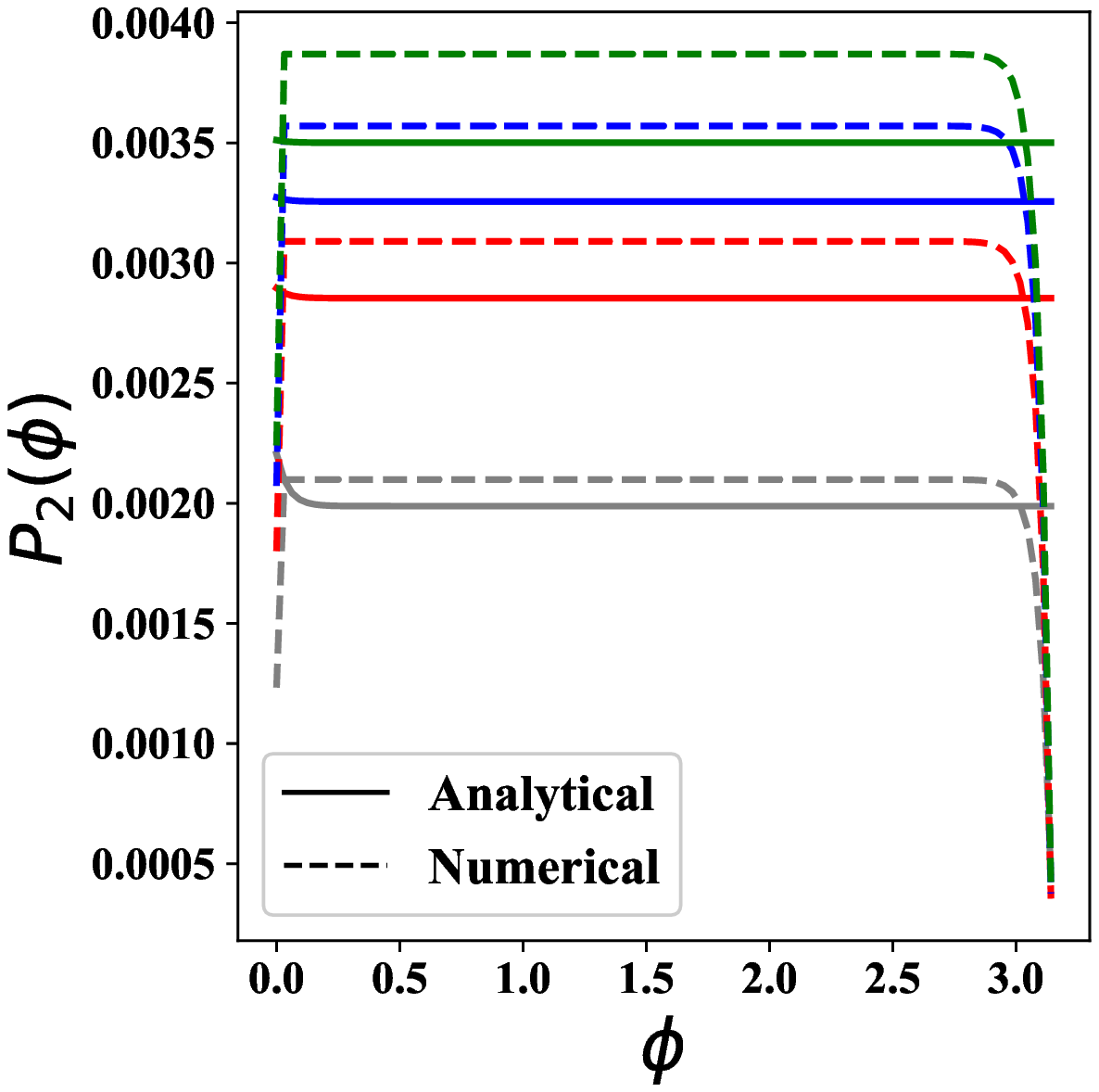}}
\def\little{\includegraphics[width=0.4\textwidth]{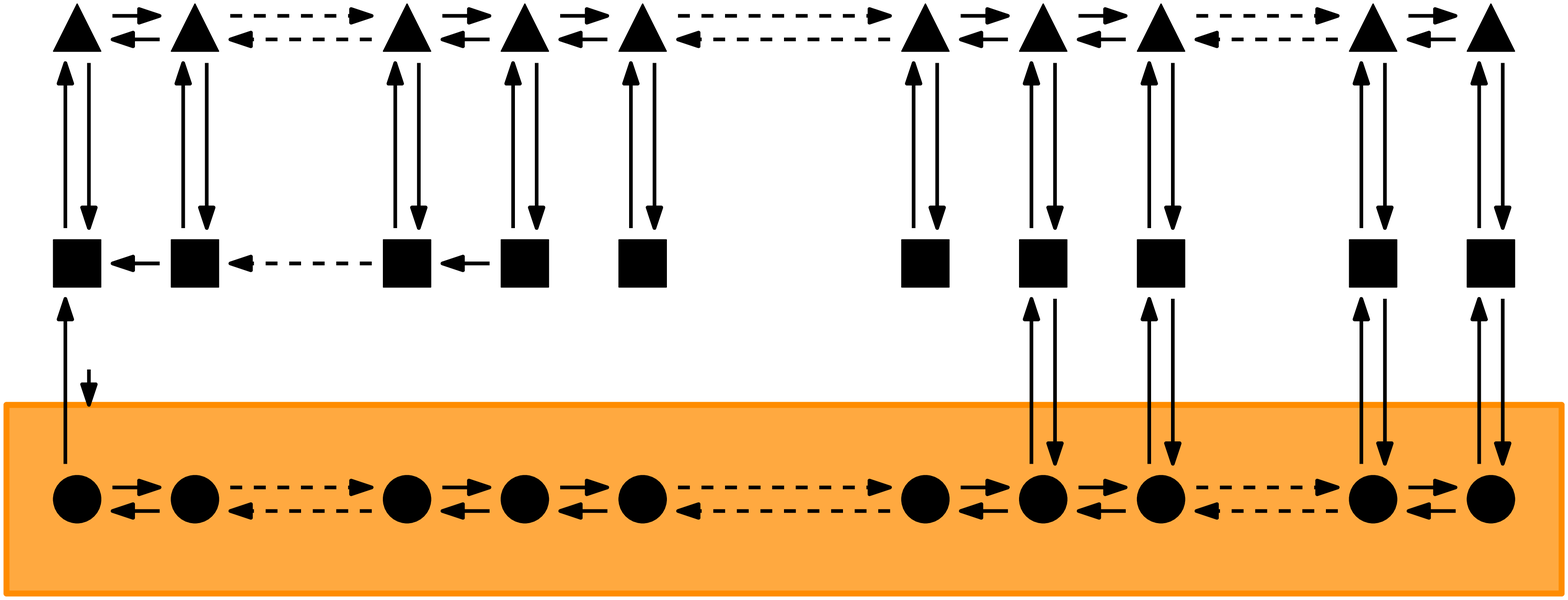}}
\stackinset{l}{0.35\textwidth}{b}{0.35\textwidth}{\little}{\big}
\caption{Comparison between numerical and analytical results for time-independent solution of $P_2$ states. The figure in inset is a representation of the Markov State network with the $P_1$ states highlighted.}
\label{P2AnalyticalNumericalComparison}
\end{subfigure}
\begin{subfigure}[b]{0.37\textwidth}
\centering
\def\big{\includegraphics[width=\textwidth]{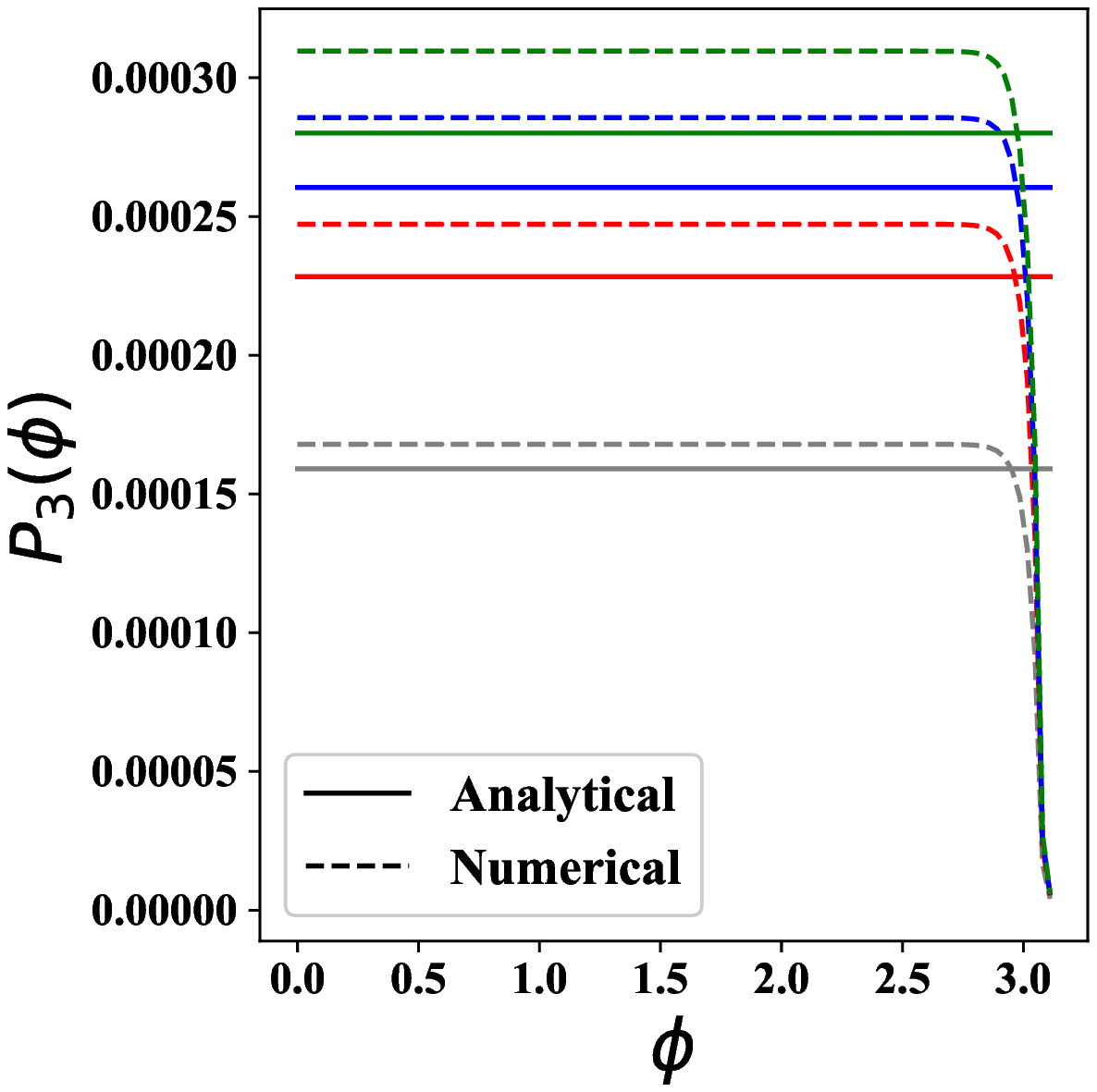}}
\def\little{\includegraphics[width=0.4\textwidth]{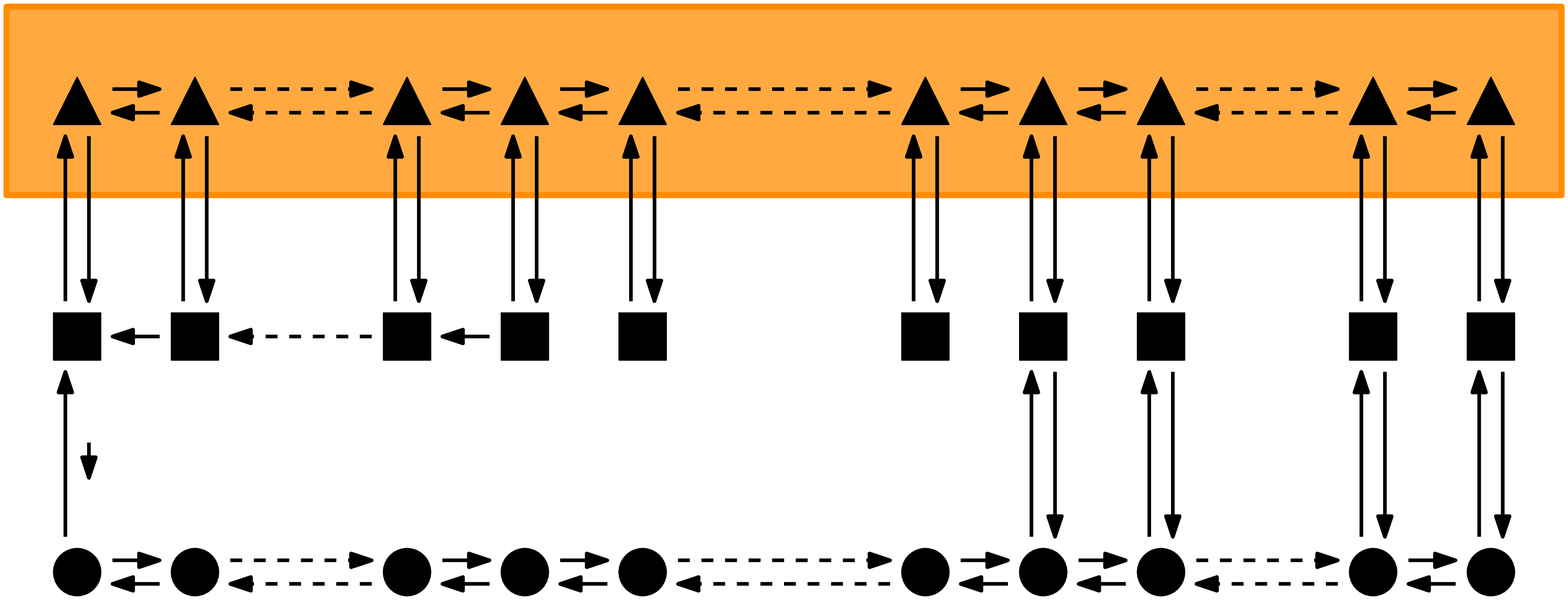}}
\stackinset{l}{0.35\textwidth}{b}{0.35\textwidth}{\little}{\big}
\caption{Comparison between numerical and analytical results for time-independent solution of $P_3$ states. The figure in inset is a representation of the Markov State network with the $P_1$ states highlighted.}
\label{P3AnalyticalNumericalComparison}
\end{subfigure}
\caption{In the main figures, \textit{grey} corresponds to $\alpha = 2$, \textit{red} to $\alpha = 4$, \textit{blue} to $\alpha = 6$ and \textit{green} to $\alpha = 8$}
\end{figure}

\begin{figure}[thb]
\begin{subfigure}[b]{0.37\textwidth}
\centering
\includegraphics[width=\textwidth]{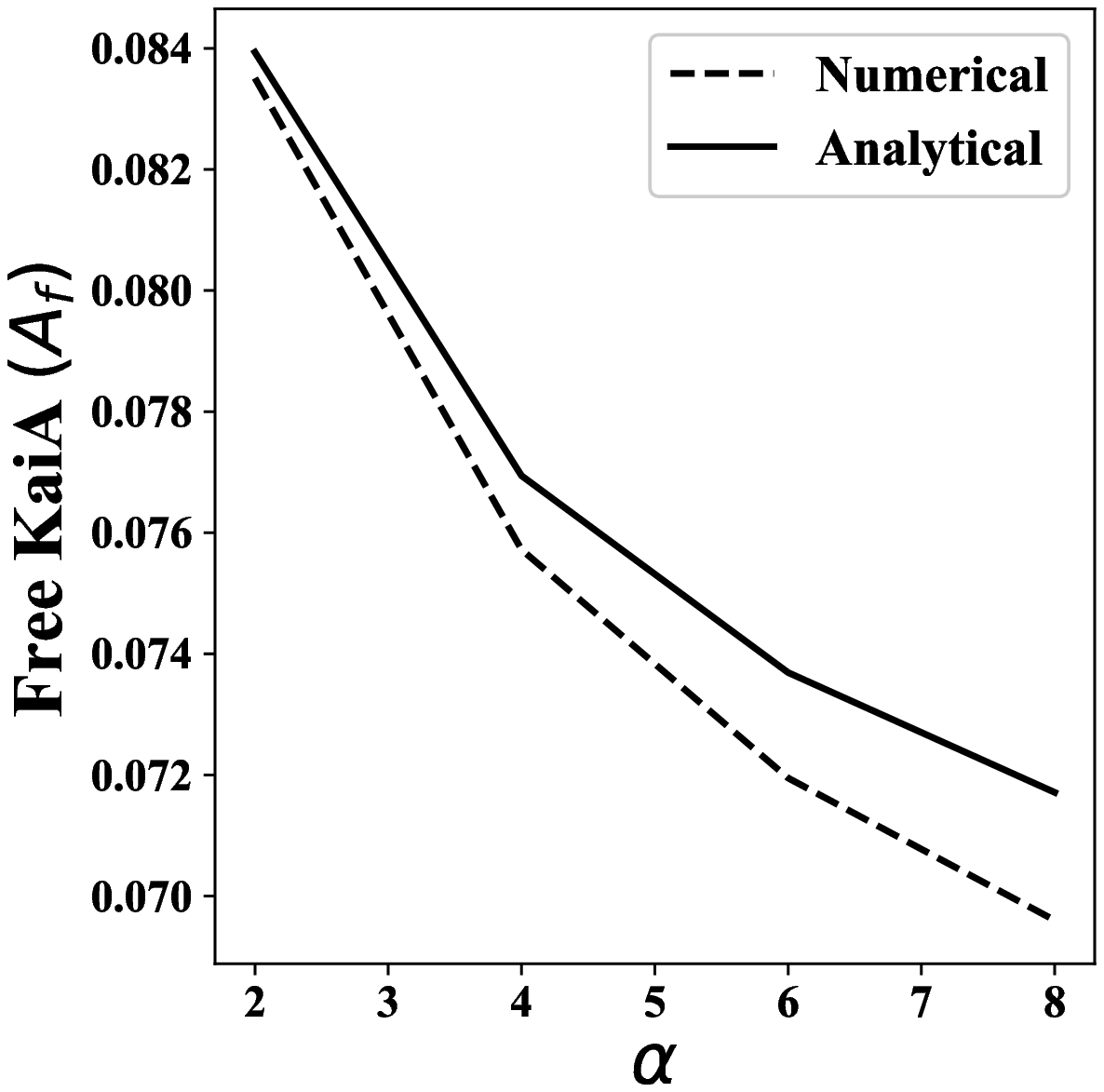}
\caption{Comparison between numerical and analytical results for free KaiA concentration.}
\label{AfAnalyticalNumericalComparison}
\end{subfigure}
\begin{subfigure}[b]{0.37\textwidth}
\centering
\includegraphics[width=\textwidth]{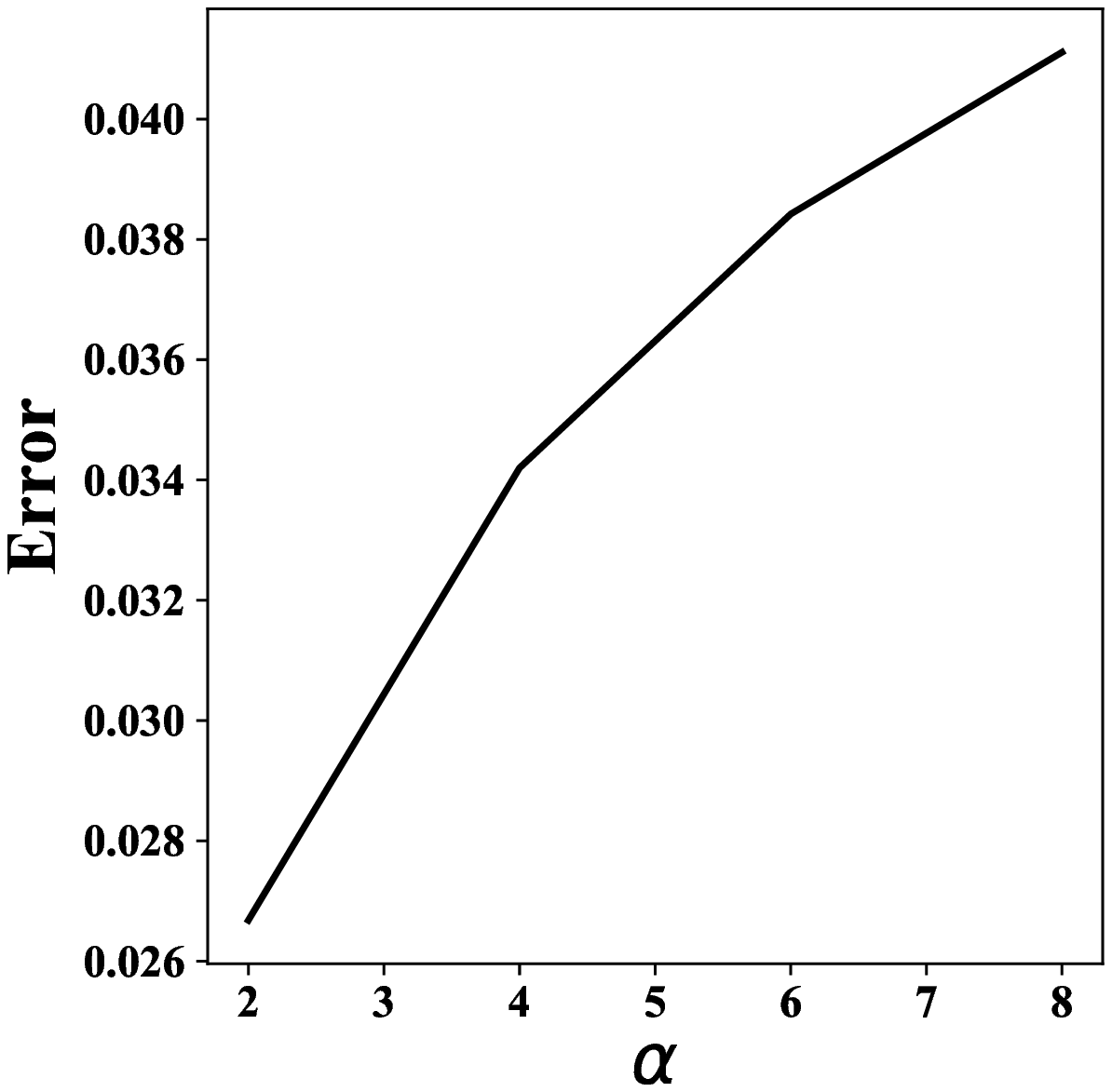}
\caption{Comparison of error between numerical and analytical results. Error = $1 - \left|\frac{\vec{P^s}_{numerical}\vec{P^s}_{analytical}}{\vec{P^s}_{numerical}\vec{P^s}_{numerical}} \right|$}
\label{ErrorAnalyticalNumericalComparison}
\end{subfigure}
\caption{Comparison of free KaiA concentration and error between the analytical data and numerical data.}
\end{figure}

\subsection{Case II : $k_1 \neq 0$}
\label{k1notzeroCalculation}
The case with $k_1 \neq 0$ is challenging to solve. Unlike the previous case, where only a single flux existed in the entire system, in this case there will be many fluxes in the system.

In order to obtain the rough form of solution for $P_1$, $P_2$ and $P_3$ states, we make some assumptions which are supported by numerical observations. We also go the continuum limit where the discrete master equations describing the system become a set of coupled PDE's. The boundaries for our problem are $x = 0$ and $x = x_0$. $x_0$ is the point where the $P_1-P_2$ connections start. Numerically it is observed that at the steady state, the probability density in the states beyond $x_0$ is negligible compared to the ones before it. Thus we set it as our boundary.  We solve the problem for the states in the bulk and then impose certain conditions such that the boundary conditions are satisfied.
\begin{align}
    \frac{\partial P_1(x)}{\partial t} &= \delta(x \neq N)k_1(P_1(x+\Delta x) - \gamma_1P_1(x)) + \delta(x \neq 0)k_1(\gamma_1P_1(x - \Delta x) - P_1(x)) \nonumber\\
	&+\delta(x=0)(\omega_1 P_2(x) - K_{d1} \omega_1 P_1(x)) \nonumber\\
	&+ H(x - x_0)(K_d \omega P_2(x) - \omega P_1(x)) -k_{Af} A_f P_1(j) + k_{Ab} \alpha^{x} P_3(j) \\
\end{align}
and so on and so forth for $P_2$ and $P_3$ states. Using Taylor expansion, $P(x+\Delta x) = P(x) + \frac{\partial P(x)}{\partial x}\Delta x + \frac{1}{2} \frac{\partial^2 P(x)}{\partial x^2} \Delta x^2 + O(\Delta x^3)$, we get,
\begin{align}
    \frac{\partial P_1(x)}{\partial t} &= k_{1c} \frac{\partial P_1(x)}{\partial x} + k_{1d} \frac{\partial^2 P_1(x)}{\partial x^2} - k_{Af} A_f P_1(x) + k_{Ab} \alpha^x P_3(x) \\
    \frac{\partial P_2(x)}{\partial t} &= k_{2c} \frac{\partial P_2(x)}{\partial x} + k_{2d} \frac{\partial^2 P_2(x)}{\partial x^2} \\
    \frac{\partial P_3(x)}{\partial t} &= k_{3c} \frac{\partial P_3(x)}{\partial x} + k_{3d} \frac{\partial^2 P_3(x)}{\partial x^2} + k_{Af} A_f P_1(x) - k_{Ab} \alpha^x P_3(x) \\
    k_{1c} &= k_1(1-\gamma_1)\Delta x, \ k_{2c} = k_{dp}(1-\gamma_2)\Delta x, \ k_{3c} = -k_0(1-\gamma)\Delta x \\
    k_{1d} &= \frac{1}{2} k_1(1+\gamma_1) \Delta x^2, \ k_{2d} = \frac{1}{2} k_{dp}(1+\gamma_2) \Delta x^2, \ k_{3d} = \frac{1}{2} k_0(1+\gamma) \Delta x^2
\end{align}

We begin with the ansatz that when $k_1$ is increased from 0 to a very small number gradually, the changes in the form of the probability distribution will not change drastically. Keeping this in mind we make the assumption, $k_{Af}A_f P_1(x) \approx k_{Ab} \alpha^x P_3(x) \forall x\in Bulk$. This assumption has been inspired by our solution for the $k_1 = 0$ case and also supported by numerical observations. It can be better written as,
\begin{align}
    P_1(x) = \frac{K_{d0}}{A_f} \alpha^x P_3(x)  
\label{approximation1:k1not0}
\end{align}
, where $K_{d0} = \frac{k_{Ab}}{k_{Af}}$. 
Thus, we have, 
\begin{align}
    \partial_x P_1(x) &= \frac{K_{d0}}{A_f} \alpha^x \left[ ln(\alpha) + \partial_x P_3(x) \right] \\
    \partial_x^2 P_1(x) &= \frac{K_{d0}}{A_f} \alpha^x \left[ (ln(\alpha))^2 P_3(x) + 2ln(\alpha) \partial_x P_3(x) + \partial_x^2 P_3(x) \right]
\end{align}

Adding the evolution equations for $P_1$ and $P_3$, in the bulk, and substituting the approximation \eqref{approximation1:k1not0} we get,
\begin{align}
\partial_t (P_1(x) + P_3(x)) &= \frac{K_{d0}}{A_f} \alpha^x ln(\alpha) [k_{1c} + k_{1d}ln(\alpha)] P_3(x) \nonumber\\
& + \left[ k_{3c} + \frac{K_{d0}}{A_f} \alpha^x (k_{1c} + k_{1d} ln(\alpha)) \right] \partial_x P_3(x) + \nonumber \\ 
& + \left[ k_{3d} + \frac{K_{d0}}{A_f} \alpha^x  k_{1d} \right] \partial_x^2 P_3(x)
\end{align}

At steady state, $\partial_t P_1(x) = 0 = \partial_t P_3(x) \ \forall x$. For $k_1 = 0$, we have $k_{1c} = 0  = k_{1d}$. Thus we have the simple ODE,
$$k_{1c} \partial_x P_3(x) + k_{3d}\partial_x^2 P_3(x) = 0$$
This can have constant solutions for $P_3(x)$ and this is exactly what we have in the case when $k_1 = 0$ \eqref{P3expression_k1_0}. The presence of $k_1$ adds an extra term dependent on $P_3$ and due to this term we cannot have constant solutions for $P_3(x)$ (unless the constant solution is $P_3(x) = 0 \forall x$). Under the assumption that the form of $P_3(x)$ does not deviate significantly from the solution when $k_1 = 0$ ($P_3(x)$ = b = constant), we can ignore terms containing $\partial_x^2 P_3(x)$. We also have $k_{3d} << k_{3c}, \ k_{1d} << k_{1c}$. Using this we can ignore terms containing $k_{1d}$ and $k_{3d}$. Thus we finally arrive at the equation,
\begin{align}
    \label{k1neq0soln}
    \frac{K_{d0}}{A_f} \alpha^x ln(\alpha) k_{1c} P_3(x) &+ \left[ k_{3c} + \frac{K_{d0}}{A_f} \alpha^x k_{1c} \right] \partial_x P_3(x) = 0 \\
    \implies P_3(x) &= P_3(x') \frac{|B + A\alpha^{x'}|}{|B + A\alpha^{x}|} \label{P3expression_k1_non0} \\
    B &= k_{3c}, \ A = \frac{K_{d0}}{A_f}k_{1c}
\end{align}

Beyond the boundary at $x = x_0$, the network is constructed in such a way that it either drives the probabilities into the $P_2$ states which are further driven towards the boundary at $x_0$ from right or it drives the probabilities towards the boundary at $x_0$ in the $P_1$ states.

Thus we can safely assume that probability of the finding a state beyond the boundary at $x_0$ is close to 0. This is confirmed by numerical results. Now we can focus our entire attention to the region, $x\in [0,x_0]$. By using conservation of flux we can find the probabilities of all the other states.

\begin{figure}[thb]
\begin{subfigure}[b]{0.4\textwidth}
\centering
\def\big{\includegraphics[width=\textwidth]{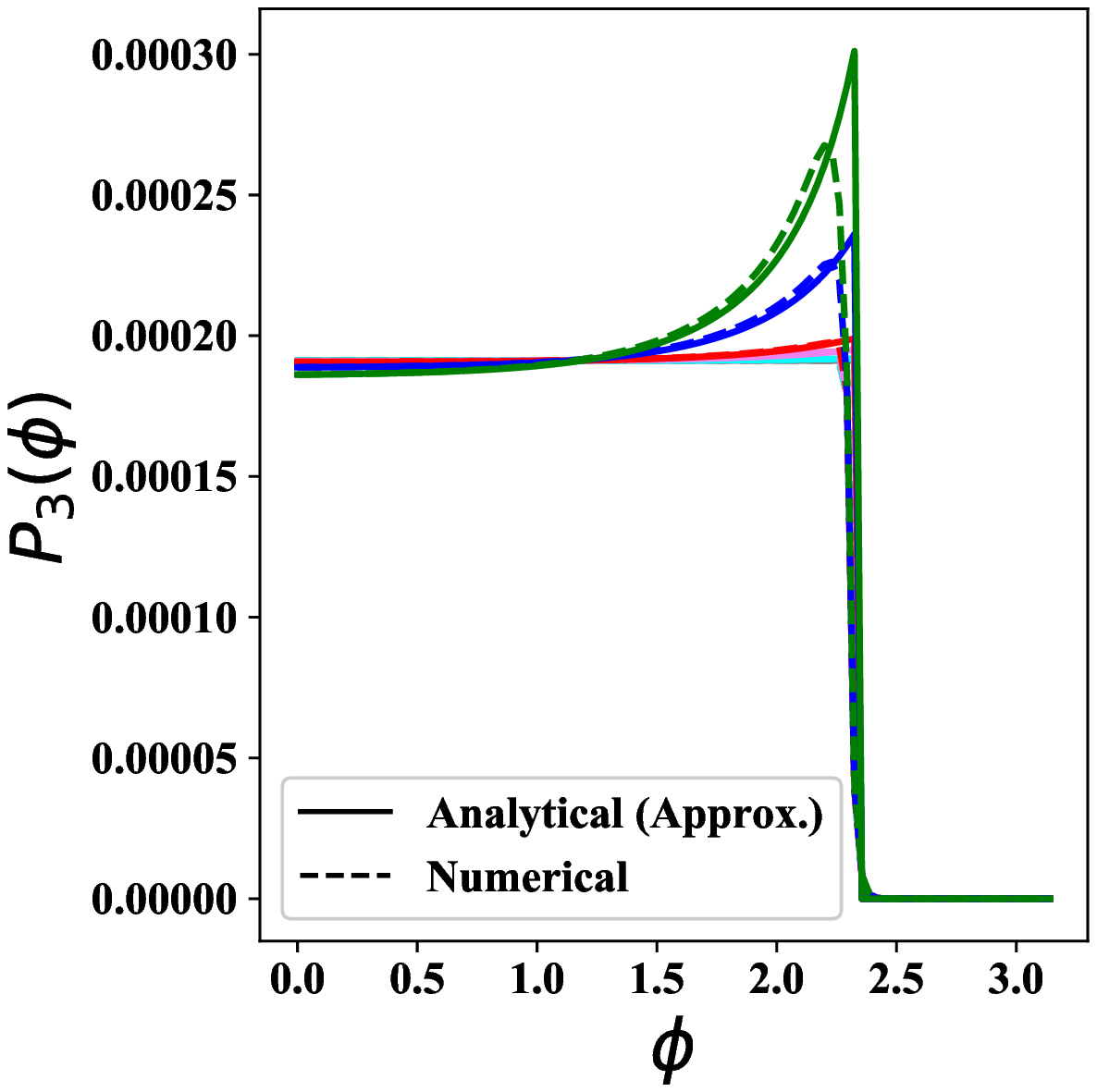}}
\def\little{\includegraphics[width=0.4\textwidth]{SIFig3b_5a_inset.eps}}
\stackinset{l}{0.35\textwidth}{b}{0.35\textwidth}{\little}{\big}
\caption{Comparison between numerical and analytical results for time-independent solution of $P_3$ states. The figure in inset is a representation of the Markov State network with the $P_3$ states highlighted.}
\label{P3AnalyticalNumericalComparison_k1_original}
\end{subfigure}
\begin{subfigure}[b]{0.4\textwidth}
\centering
\def\big{\includegraphics[width=\textwidth]{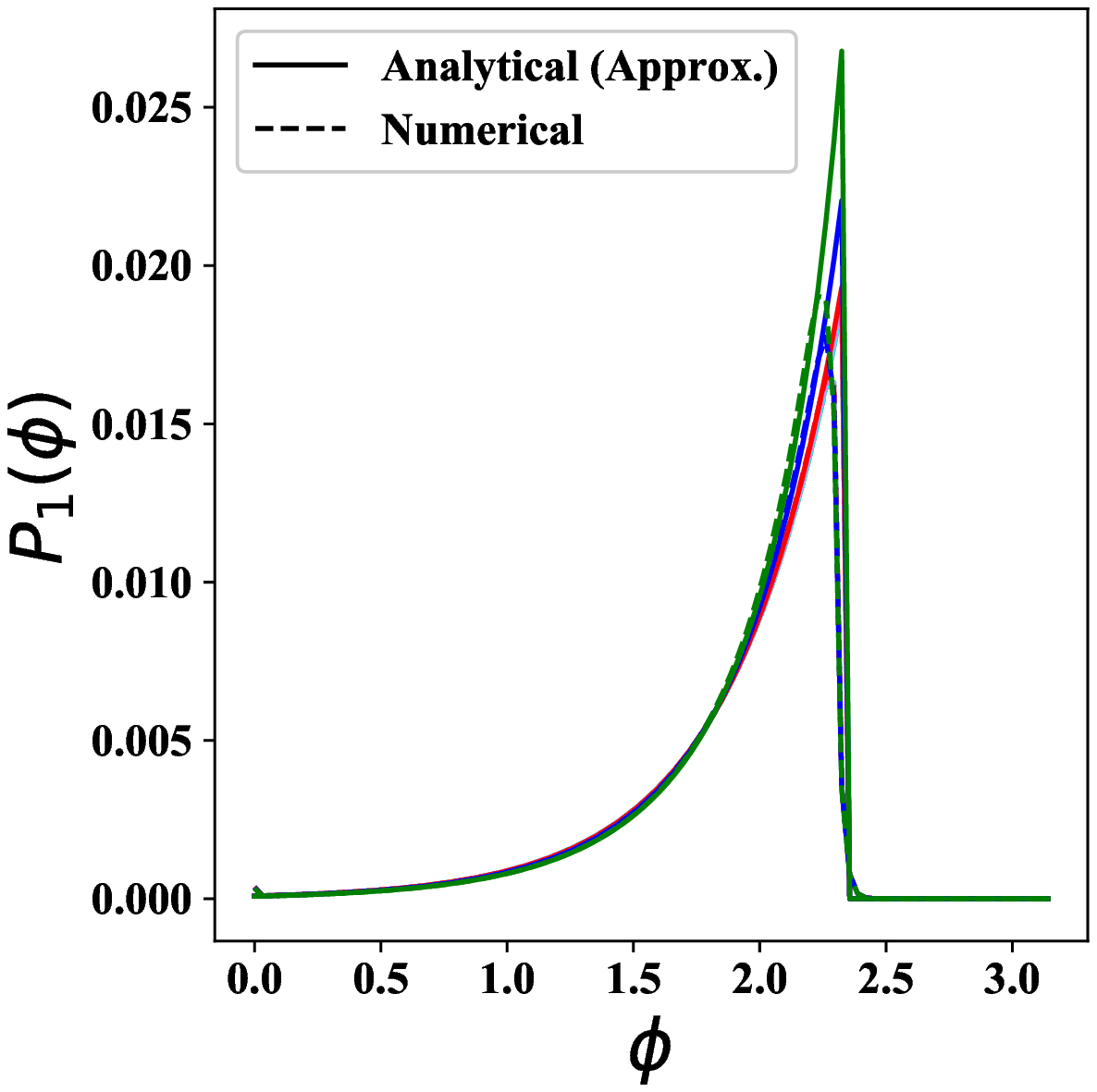}}
\def\little{\includegraphics[width=0.4\textwidth]{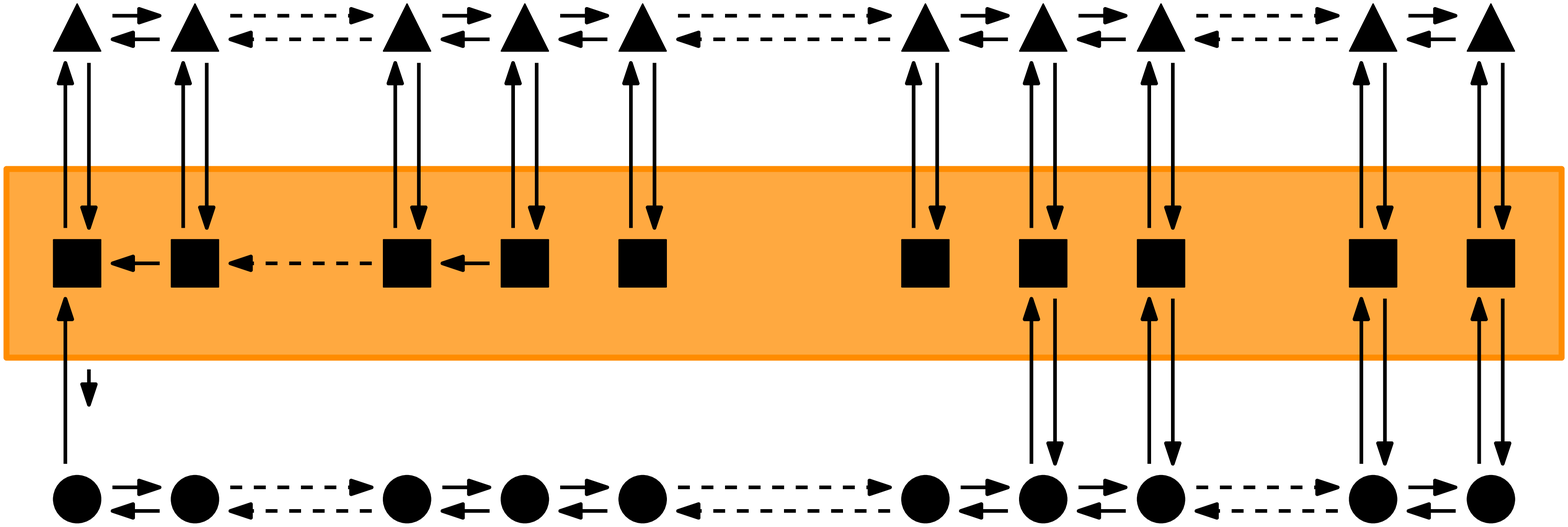}}
\stackinset{l}{0.25\textwidth}{b}{0.5\textwidth}{\little}{\big}
\caption{Comparison between numerical and analytical results for time-independent solution of $P_1$ states. The figure in inset is a representation of the Markov State network with the $P_1$ states highlighted.}
\label{P1AnalyticalNumericalComparison_k1}
\end{subfigure}
\caption{In the main figures, \textit{grey} corresponds to $k_1 = 0$, \textit{cyan} to $k_1 = 10^{-4}$, \textit{violet} to $k_1 = 5\times10^{-4}$, \textit{red} to $k_1 = 10^{-3}$, \textit{blue} to $k_1 = 5\times10^{-3}$, \textit{green} to $k_1 = 10^{-2}$.}
\end{figure}

\subsection{Calculating Time-Independent Steady State numerically}
As mentioned earlier, when $k_1 \neq 0$ and there are more than one connections between $P_1$ and $P_2$ states i.e. $j_0, x_0 \neq N$, we have multiple fluxes in the system. Nevertheless, we can still find the steady-state time-independent solution for $\vec{P}$ irrespective of whether it is stable or not. An iterative procedure is adopted. The first step in this procedure is to find the free KaiA concentration in the system. In the following paragraph the procedure is described.
The set of FPE's that describe the evolution of $\vec{P}$ can be expressed as, $\frac{\partial \vec{P}}{\partial t} = \boldmath{W(\vec{P})}\vec{P}$, where $\vec{P}$ is a vector of length $3N+3$. The first N+1 elements would correspond to $P_1$ form, the next N+1 elements would correspond to $P_2$ form and the last N+1 elements would correspond to $P_3$ form. The rate matrix, W is function of the probabilities due to the presence of the $A_f$ term which makes the entire thing non-linear. Now if we succeed in finding the free KaiA concentration at steady state, then substituting it back into W would make it a linear system to solve, and then $\frac{\partial \vec{P}}{\partial t} = \boldmath{W}\vec{P}$. We can find $A_f^s$ (free KaiA at steady state) using an iterative procedure as follows:
\begin{enumerate}
	\item Initialize $A_f = A_t$ for the first run and form the rate matrix, W.
	\item At steady state, $\frac{\partial \vec{P}}{\partial t} = 0 = W\vec{P}$. Compute the eigenvector corresponding to the nullspace of W and call it $\vec{v_0}$.
	\item Compute $A_f^{test} = A_t - \epsilon \sum_{i=2N+2}^{3N+2} v_0(i) - \epsilon_{seq}\sum_{i=N+1}^{2N+1}v_0(i) $
	\item If $A_f^{test} < 0$, it would be unphysical. So set, $A_f^{new} = \frac{A_f^{old}}{2}$, else set $A_f^{new} = A_f^{old} + \delta(A_f^{test}-A_f^{old})$ where $\delta$ is some appropriate step size.
	\item Repeat this procedure until convergence i.e. $\frac{|A_f^{old} - A_f^{test}|}{A_f^{old}} < Tolerance$
\end{enumerate}
Once we have $A_f^s$ we can find $\vec{P^s}$.

\section{Linear Stability Analysis}
\label{LinearStabilityAnalysis}

We perturb around the steady state distribution, $\vec{P^s}$. Say, $P_k(j) = P_k^s(j) + \delta \eta_k(j)$, $k = 1,2,3$ and $j = {0,...,N}$. By conservation of probability, we have $\sum_{k,j} \eta_k(j) = 0$. Substituting $P_k(j)$ in the differential equations, lead us to the evolution equations for $\eta_k(j)$.

\begin{align}
	\frac{\partial \eta_1(j)}{\partial t} &= \delta(j \neq N)k_1(\eta_1(j+1) - \gamma_1\eta_1(j)) + \delta(j \neq 0)k_1(\gamma_1\eta_1(j - 1) - \eta_1(j)) \nonumber\\
	&+\delta(j=0)(\omega_1 \eta_2(j) - K_{d1} \omega_1 \eta_1(j)) \nonumber\\
	&+ H(j - j_0)(K_d \omega \eta_2(j) - \omega \eta_1(j)) -k_{Af} A_f^s \eta_1(j) + k_{Ab} \alpha^{j} \eta_3(j)  \nonumber\\
	&+\epsilon k_{Af} P_1^s(j) \sum_{i=0}^N \eta_3(i) + \epsilon_{seq} k_{Af} P_1^s(j) \sum_{i=0}^N \eta_2(i) + O(\delta)\\
	\frac{\partial \eta_2(j)}{\partial t} &= \delta(j \neq 0)k_{dp}(\gamma_2\eta_2(j-1) - \eta_2(j)) + \delta(j \neq N)k_{dp}(\eta_2(j + 1) - \gamma_2\eta_2(j)) \nonumber\\
	&-\delta(j=0)(\omega_1 \eta_2(j) - K_{d1} \omega \eta_1(j)) \nonumber\\
	&+H(j - j_0)(\omega \eta_1(j) - \omega \eta_2(j)) \\
	\frac{\partial \eta_3(j)}{\partial t} &= \delta(j \neq N)k_{0}(\gamma \eta_3(j+1) - \eta_3(j)) + \delta(j \neq 0)k_{0}(\eta_3(j - 1) - \gamma \eta_3(j)) \nonumber\\
	&+k_{Af} A_f^s \eta_1(j) - k_{Ab} \alpha^{j} \eta_3(j) \nonumber\\
	&-\epsilon k_{Af} P_1^s(j) \sum_{i=0}^N \eta_3(i) - \epsilon_{seq} k_{Af} P_1^s(j) \sum_{i=0}^N \eta_2(i) + O(\delta)
\end{align}

Notice the additional terms in evolution of $\vec{\eta_1}$ and $\vec{\eta_3}$ which are directly dependent on $\epsilon$ and $\epsilon_{seq}$.
The entire thing can be expressed as $\frac{\partial \vec{\eta}}{\partial t} = \tilde{\mathbf{W}} \vec{\eta} = (\mathbf{W} + \mathbf{W'}) \vec{\eta}$. The entire matrix $\tilde{W}$ can be broken into 9 parts, each representing interactions between different types of states as shown in \eqref{Wmatrix}. The interesting blocks in the W-matrix are the $\eta_3-\eta_3$ \eqref{eta3eta3block} and $\eta_3-\eta_2$ \eqref{eta3eta2block} blocks which contain most of the terms arising due to nonlinearities.

In short, a linear stability analysis can be performed around the steady state of the system, $\vec{P^s}$, which yields upto first order,
\begin{align}
    \frac{\partial \vec{\eta}}{\partial t} &= [\mathbf{W(\vec{P^s})} + \nabla_{\vec{P}}\mathbf{W(\vec{P})}\vec{P}|_{\vec{P^s}}]\vec{\eta} = [\mathbf{W(\vec{P^s})} + \mathbf{W'(\vec{P^s})}] \vec{\eta} = \mathbf{\tilde{W}} \vec{\eta}
    \label{UptoFirstOrder}
\end{align}
where $\eta$ is the vector of perturbation (see Section \ref{LinearStabilityAnalysis} for a detailed expression). We work in a regime where the \%ATP i.e. $K_{d0}$ in our model plays a major role in deciding whether oscillations take place or not. 
The initial condition for every simulation is set to $P_1(0) = 1$ at t=0. This corresponds to starting all reactions with all the KaiC in the unphosphorylated and ADP bound form. The simulations are allowed to run for some time in order to reach either a time-independent or a time-dependent steady state behaviour. Now let us look at the $\mathbf{W}$ matrix. It can be broken into 9 blocks,

\begin{equation}
\tilde{W} =  \left(
\begin{array}{ *{1}{c} | *{1}{c} | *{1}{c} }
   W_{11} & W_{12} & W_{13} \\\hline
   W_{21} & W_{22} & W_{23} \\\hline
   W_{31} & W_{32} & W_{33} 
  \end{array}
\right)
\label{Wmatrix}
\end{equation}

\begin{equation}
W_{33} =  \left(
\begin{array}{ *{1}{c}  *{1}{c}  *{1}{c} *{1}{c} *{1}{c} }
   -k_{Ab} \alpha^0 - \epsilon k_{Af} P_1^s(0)  & -\epsilon k_{Af} P_1^s(0) & -\epsilon k_{Af} P_1^s(0) & \ldots & -\epsilon k_{Af} P_1^s(0) \\
   - \epsilon k_{Af} P_1^s(1)  & -k_{Ab} \alpha^1 -\epsilon k_{Af} P_1^s(1) & -\epsilon k_{Af} P_1^s(1) & \ldots & -\epsilon k_{Af} P_1^s(1) \\
   \vdots  & \ldots & \ddots & \ldots & \vdots \\
   - \epsilon k_{Af} P_1^s(N-1)  & \ldots & -\epsilon k_{Af} P_1^s(N-1) & -k_{Ab} \alpha^{N-1} -\epsilon k_{Af} P_1^s(N-1) & -\epsilon k_{Af} P_1^s(N-1) \\
   - \epsilon k_{Af} P_1^s(N)  & \ldots & -\epsilon k_{Af} P_1^s(N) & -\epsilon k_{Af} P_1^s(N) & -k_{Ab} \alpha^N -\epsilon k_{Af} P_1^s(N) 
  \end{array}
\right)
\label{eta3eta3block}
\end{equation}

\begin{equation}
W_{32} =  \left(
\begin{array}{ *{1}{c}  *{1}{c}  *{1}{c} *{1}{c} *{1}{c} }
   - \epsilon_{seq} k_{Af} P_1^s(0)  & -\epsilon_{seq} k_{Af} P_1^s(0) & -\epsilon_{seq} k_{Af} P_1^s(0) & \ldots & -\epsilon_{seq} k_{Af} P_1^s(0) \\
   - \epsilon_{seq} k_{Af} P_1^s(1)  & -\epsilon_{seq} k_{Af} P_1^s(1) & -\epsilon_{seq} k_{Af} P_1^s(1) & \ldots & -\epsilon_{seq} k_{Af} P_1^s(1) \\
   \vdots  & \ldots & \ddots & \ldots & \vdots \\
   - \epsilon_{seq} k_{Af} P_1^s(N-1)  & \ldots & -\epsilon_{seq} k_{Af} P_1^s(N-1) & -\epsilon_{seq} k_{Af} P_1^s(N-1) & -\epsilon_{seq} k_{Af} P_1^s(N-1) \\
   - \epsilon_{seq} k_{Af} P_1^s(N)  & \ldots & -\epsilon_{seq} k_{Af} P_1^s(N) & -\epsilon_{seq} k_{Af} P_1^s(N) & -\epsilon_{seq} k_{Af} P_1^s(N) 
  \end{array}
\right)
\label{eta3eta2block}
\end{equation}

To make things simpler let us look at just the $\mathbf{W'}$ matrix elements,
\begin{align}
    \mathbf{W'} &=  \left(
    \begin{array}{ *{1}{c} | *{1}{c} | *{1}{c} }
   W'_{11} & W'_{12} & W'_{13} \\\hline
   W'_{21} & W'_{22} & W'_{23} \\\hline
   W'_{31} & W'_{32} & W'_{33} 
  \end{array}
    \right) \\
    W'_{33} &=  \left(
    \begin{array}{ *{1}{c}  *{1}{c}  *{1}{c} *{1}{c} }
       - \epsilon k_{Af} P_1^s(0)  & -\epsilon k_{Af} P_1^s(0) & \ldots & -\epsilon k_{Af} P_1^s(0) \\
       - \epsilon k_{Af} P_1^s(1)  & -\epsilon k_{Af} P_1^s(1) & \ldots & -\epsilon k_{Af} P_1^s(1) \\
       \vdots  & \ldots & \ddots & \vdots \\
       - \epsilon k_{Af} P_1^s(N)  & \ldots & -\epsilon k_{Af} P_1^s(N) & -\epsilon k_{Af} P_1^s(N) 
      \end{array}
    \right) \\
    W'_{32} &=  \left(
    \begin{array}{ *{1}{c}  *{1}{c}  *{1}{c} *{1}{c} }
       - \epsilon_{seq} k_{Af} P_1^s(0)  & -\epsilon_{seq} k_{Af} P_1^s(0) & \ldots & -\epsilon_{seq} k_{Af} P_1^s(0) \\
       - \epsilon_{seq} k_{Af} P_1^s(1)  & -\epsilon_{seq} k_{Af} P_1^s(1) & \ldots & -\epsilon_{seq} k_{Af} P_1^s(1) \\
       \vdots  & \ldots & \ddots  & \vdots \\
       - \epsilon_{seq} k_{Af} P_1^s(N)  & \ldots & -\epsilon_{seq} k_{Af} P_1^s(N) & -\epsilon_{seq} k_{Af} P_1^s(N) 
      \end{array}
    \right)
\end{align}

Calculating the eigenvalues of W tells us about the stability of the steady state. Presence of +ve eigenvalues would indicate that the steady state is unstable and that a time-dependent steady state is present in the system.This would give rise to oscillations.

\subsection{Origin of Instability}
\label{OriginOfInstability}
Increasing $\alpha$ leads to accumulation of probability density near the higher phosphorylated region of $P_1$ and $P_3$ states. Eventually, this leads to an instability. The oscillatory state is stable because higher $\alpha$ provides coherence to the wavepacket i.e. the phosphorylation wavepacket has a narrow width as it moves across the different states \cite{Zhang2020}.

One simple way to understand the emergence of oscillations with increasing $k_1$ is through the Gershgorin circle theorem. The Gershgorin circle theorem provides us a way to estimate the location of the eigenvalues of any square matrix. Simply put, it says that the for any square matrix W, if we construct the pair $(W_{ii}, R_i)$, where $R_i = \sum_{j, j\neq i} |W_{ji}|$, then all the eigenvalues of W lie in the union of circles with radii $R_i$ and centred at $W_{ii}$. The onset of instability means the presence of +ve eigenvalues in the W matrix as mentioned before. In a normal rate matrix, all the diagonal entries are -ve and the off-diagonal entries are +ve in a way such that sum of all element in each column is 0. Gershgorin theorem can be easily applied to this system and it can be seen that the eigenvalues will always have to be -ve (or 0). But in our case, the matrix W does have -ve off-diagonal elements, for instance $-\epsilon k_{Af} P_1^s(j) \eta_3(i) , i\neq j$ term in the evolution of $\eta_3(j)$. All such terms which are -ve in the off-diagonal position have $P_1^s$. It is the presence of these terms which extend the Gershgorin circles into the positive half of the plane. So, our chances of obtaining a positive eigenvalue increases if we have higher $P_1^s(j) \forall j$. Now it remains to show that as $K_{d0}$ increases, $P_1^s$ either decreases or stays unchanged and when $k_1$ increases, $P_1^s$ increases.

From \eqref{k1neq0soln} we know the forms of the solution for $k_1 \neq 0$. For a moment let us take $A_f$ to be fixed. This assumption is justified in the limit when $k_1$ is very small and there is not much change in the value of $A_f$ derived in the $k_1 = 0$ case. For this fixed value of $A_f$ let us consider two cases, the first when $k_1 = 0$ and the second when $k_1 \neq 0$. Fixed $A_f$ implies that $\sum P_3(i) = constant = c$. Let us denote the functional form for $P_3(x)$ as f(x) for case I and g(x) for case II. As we have shown previously f(x) is a constant function and g(x) is a strictly increasing function. $\sum_x (f(x) - g(x)) = 0$ and the fact that g(x) is strictly increasing implies that f(x) and g(x) have a single point where they cross each other i.e. $f(x)>g(x)$ for $x<x_c$, $f(x)=g(x)$ at $x=x_c$ and $f(x)<g(x)$ for $x>x_c$, where $x_c$ is the point of crossover. Now $\sum_x(P_1^{II}(x) - P_1^{I}(x)) = \frac{K_{d0}}{A_f}\sum_x \alpha^x(g(x) - f(x))$. This is a polynomial in $\alpha$ with a single sign change in the coefficients at $x = x_c$, with 1 as a root and with the leading term $(g(x_N)-f(x_N))$ to be positive. Thus from Decartes rule for change in signs we can can say that 1 is the only positive root of the polynomial and thus $\sum_x \alpha^x(g(x) - f(x)) > 0 \ \forall \alpha>1$. Thus we can say that increasing $k_1$ increases $\sum_{x} P_1(x)$. This in turn affects the radii of the Gershgorin circles and thus the possibility of having an eigenvalue in the positive half of the complex plane increases with increasing $k_1$.

\begin{figure}[th]
    \label{P1ssVsk1}
	\centering
	\begin{subfigure}[b]{0.45\textwidth}
	\includegraphics[width = \textwidth]{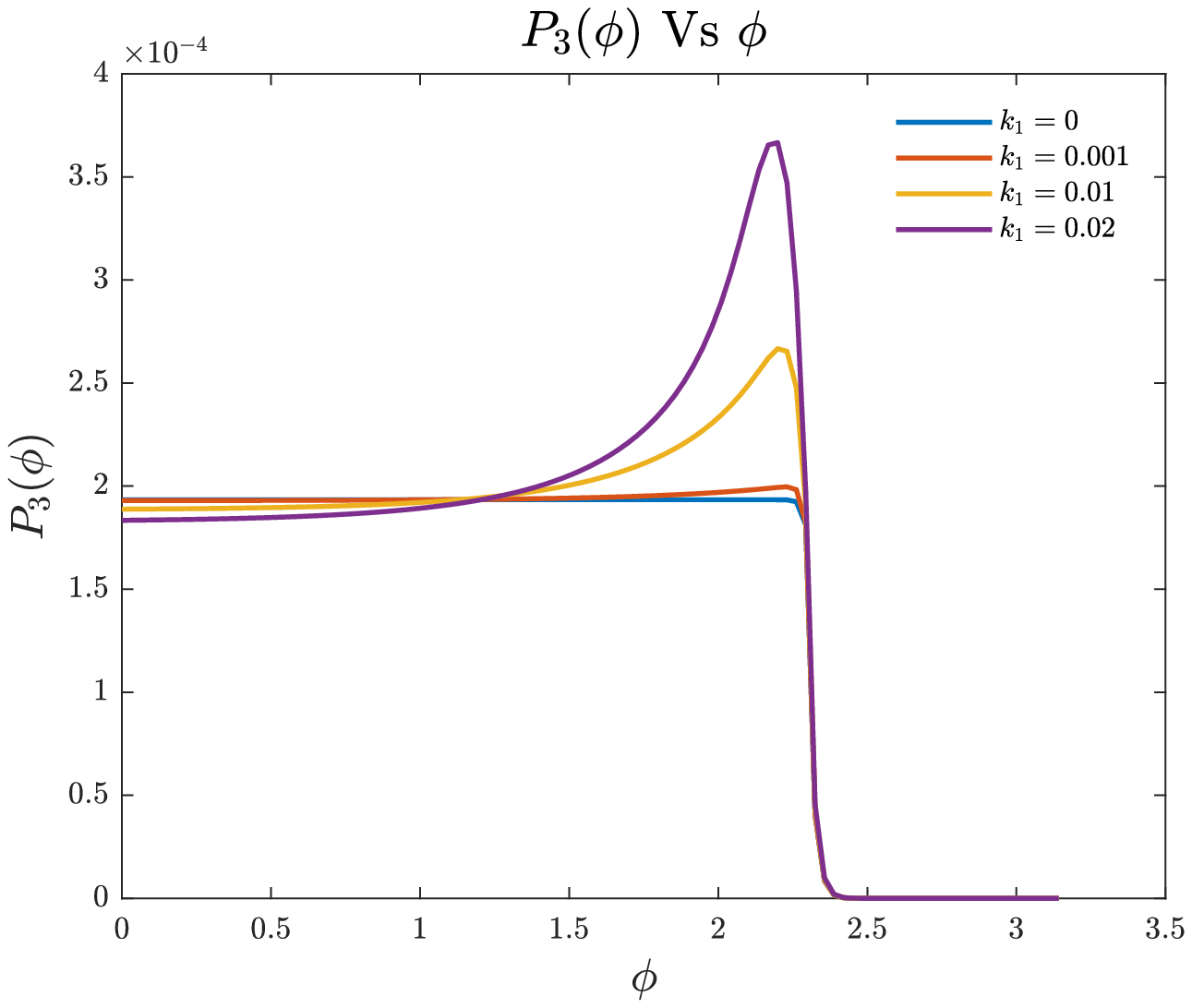}
	\caption{For $k_1=0$ we have a constant solution for $P_3(\phi)$. For $k_1\neq0$, we have a solution of the form \eqref{P3expression_k1_non0}.}
	\end{subfigure}
	\begin{subfigure}[b]{0.45\textwidth}
	\includegraphics[width = \textwidth]{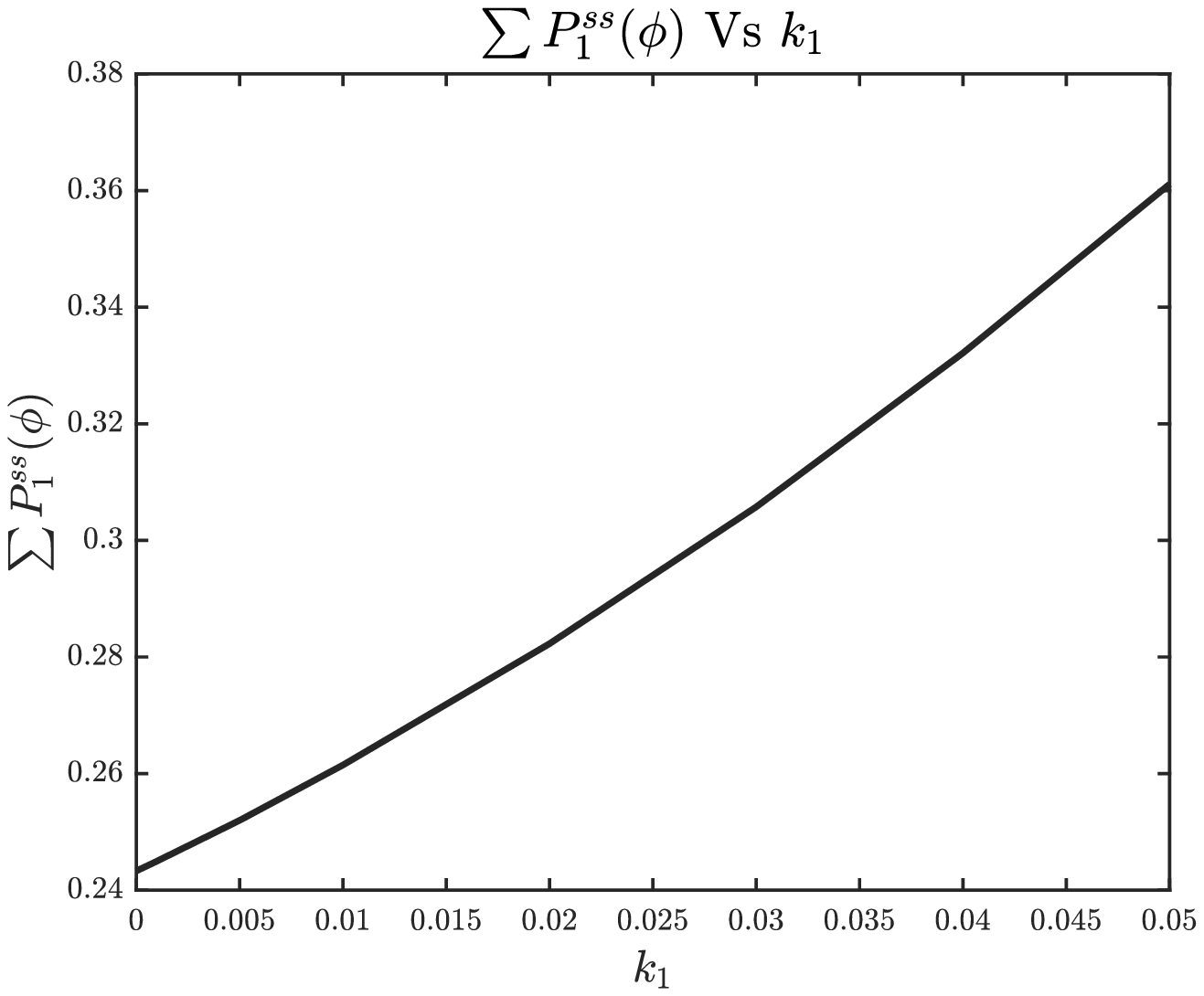}
	\caption{This graph shows how the total probability in the $P_1$ form at steady state increases as a function of $k_1$.}
	\end{subfigure}
	\caption{Origin of Instability}
\end{figure}


\begin{figure}[ht]
	\centering
	\includegraphics[width = 0.8\textwidth]{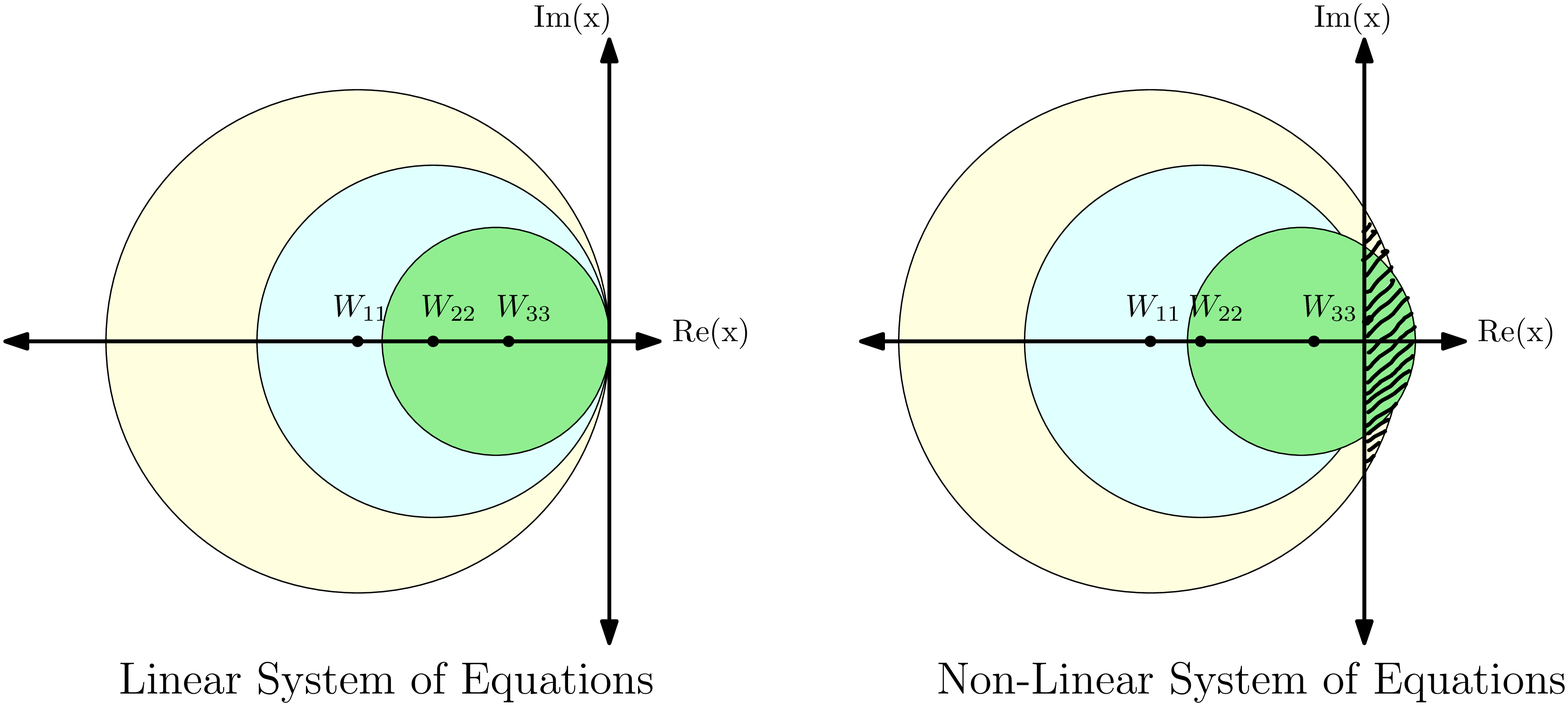}
	\caption{All the eigenvalues lie in within the area of the union of all circles. In The linear case, the union of all circles lies completely in the left half of the Argand plane. Thus no eigenvalues are possible which have a +ve real part. In the nonlinear case, due to the -ve off diagonal terms, there is a possibility to have eigenvalues with +ve real part. If the shaded reagion has a greater area then chances of getting an instability increases.}
	\label{Greshgorin}
\end{figure}

\section{First Passage analysis}
\label{FirstPassageTime}

\begin{figure}[H]
\centering
\includegraphics[width = 0.8\textwidth]{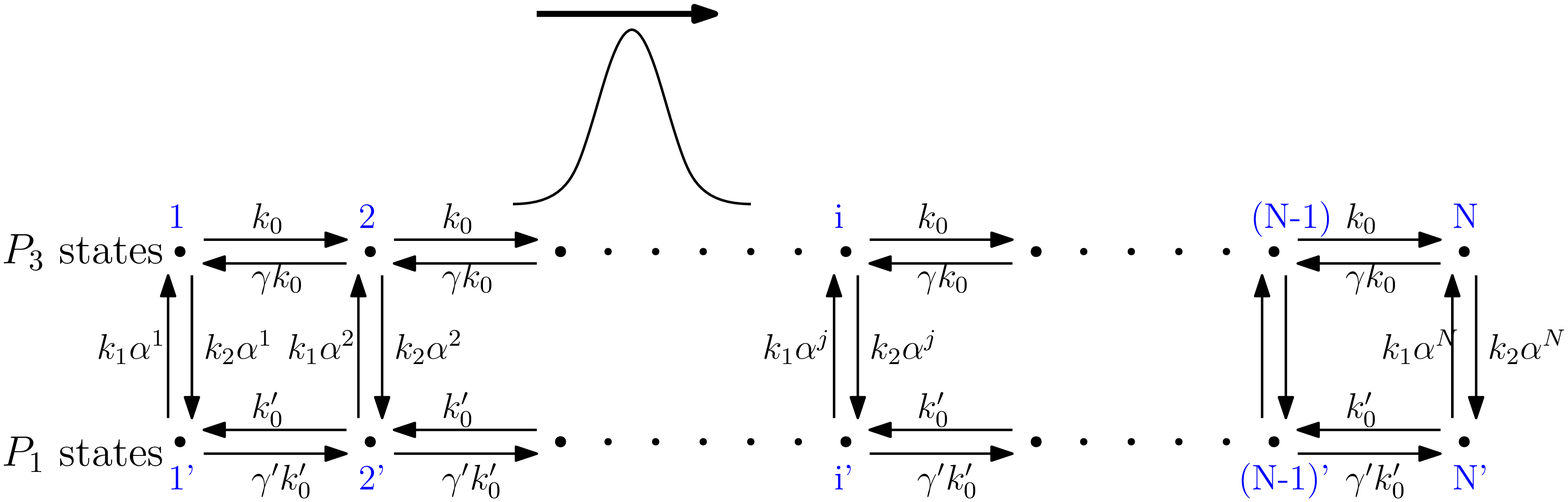}
\caption{The wave of phosphorylation moves across the $P_1$ and $P_3$ states. The movement of the wavepacket can be realized as a particle hopping between sites labelled with indices $i$ (for $P_3$ states) and $i'$ (for $P_1$ states)} 
\end{figure}

In the notation used to solve the Fokker Planck equations, $k_1 \alpha^x$ would correspond to $k_{A_f} A_f$, $k_2 \alpha^x$ would correspond to $k_{Ab0} \alpha^{x - \pi}$, $k_0'$ to $k_1$ and $\gamma_0'$ to $\gamma_1$. It is observed as the wavepacket moves, the free KaiA concentration changes with time as a function of $\alpha$ as, $A_f = A_{f0} \alpha^x$ where x is the position of the tip of the wavepacket.
 
Let $T_n$ denote the time taken for the particle to reach the end for the first time, starting from the $n^{th}$ position.
\begin{align}
    T_1 &= \frac{k_2 \alpha}{k_2 \alpha + k_0} \left( T_{1'} + \frac{1}{k_2 \alpha} \right) + \frac{k_0}{k_0 + k_2 \alpha} \left( T_2 + \frac{1}{k_0} \right) \\
    T_{1'} &= \frac{k_1 \alpha}{k_1 \alpha + \gamma' k_0'} \left( T_1 + \frac{1}{k_1 \alpha} \right) + \frac{\gamma' k_0'}{k_1 \alpha + \gamma' k_0'}\left( T_2' + \frac{1}{\gamma' k_0'} \right) \\
    T_2 &= \frac{k_2 \alpha^2}{k_0 (1+\gamma) + k_2 \alpha^2}\left( T_{2'} + \frac{1}{k_2 \alpha^2}\right) + \frac{k_0}{k_0(1+\gamma) + k_2 \alpha^2} \left( T_3 + \frac{1}{k_0} \right) + \frac{\gamma k_0}{k_0(1+\gamma) + k_2 \alpha^2} \left( T_1 + \frac{1}{\gamma k_0} \right) \\
    T_{2'} &= \frac{k_1 \alpha^2}{k_0' (1+\gamma') + k_1 \alpha^2}\left( T_{2} + \frac{1}{k_1 \alpha^2}\right) + \frac{\gamma' k_0'}{k_0'(1+\gamma') + k_1 \alpha^2} \left( T_{3'} + \frac{1}{\gamma' k_0'} \right) + \frac{k_0'}{k_0'(1+\gamma') + k_1 \alpha^2} \left( T_{1'} + \frac{1}{k_0'} \right)
\end{align}
and so on and so forth.
Define:
\begin{align}
    s_j = k_0(1+\gamma) + k_2 \alpha^j \, \ p_j = \frac{k_2 \alpha^j}{s_j} \, \ q_j = \frac{k_0}{s_j} \, \ r_j = \frac{\gamma k_0}{s_j} 
\end{align}
and similarly for the indices $j'$. Using these definitions, we have,
\begin{align}
    T_j &= p_j T_{j'} + q_j T_{j+1} + r_j T_{j-1} + 3s_j^{-1} \\
    T_{j'} &= p_{j'} T_j + r_{j'} T_{j' + 1} + q_{j'} T_{j' - 1} + 3s_j^{-1} \\
    \implies T_j &= T_{j'} + \frac{3}{p_j s_j} + \frac{q_j - r_j}{p_j} \frac{\partial T_j}{\partial j} \\
    T_{j'} &= T_j + \frac{3}{p_{j'} s_{j'}} + \frac{r_j - q_j}{p_j} \frac{\partial T_{j'}}{\partial j'}
\end{align}
Eliminating T' and relabelling j by x, we get,
\begin{align}
    \frac{(1-\gamma)(1 - \gamma') k_0 k_0'}{k_1 k_2 \alpha^x} \frac{\partial^2 T}{\partial x^2} + \left[ \frac{(1-\gamma) k_0}{k_2} -\frac{(1-\gamma')k_0'}{k_1} \right] \frac{\partial T}{\partial x} + 3\left[ \frac{1}{k_1} + \frac{1}{k_2} - \frac{(1- \gamma')k_0' ln\alpha}{k_1 k_2 \alpha^x} \right] = 0
\end{align}
Assuming that T is linear in x for small $k_0'$, the first term can be neglected. The solution of T(x) is given as,
\begin{align}
    A T(x) &= -\frac{C}{ln\alpha} (1 - \alpha^{-L}) - B(x - L) \\
\end{align}
where, $A = \frac{(1-\gamma) k_0}{k_2} -\frac{(1-\gamma')k_0'}{k_1}$, $B = \frac{1}{k_1} + \frac{1}{k_2}$ and $C = \frac{(1-\gamma')k_0' ln\alpha}{k_1 k_2}$
The velocity of the wave packet is given by, $v = \frac{L}{T(0)}$. Putting in all the values, we get,
\begin{align}
    v = \frac{1}{3} \frac{(1-\gamma)k_0 - \frac{k_2}{k_1}(1 - \gamma')k_0'}{1 + \frac{k_2}{k_1} - \frac{k_0'}{k_1}(1 - \gamma')ln\alpha} \approx \frac{1}{3} \frac{(1-\gamma)k_0 - \frac{k_2}{k_1}(1 - \gamma')k_0'}{1 + \frac{k_2}{k_1}}
\end{align}
In the regime where we have oscillations, $k_0'<<k_1$. Thus the second term in the denominator can be ignored. We have, $\frac{k_2}{k_1} = \frac{k_{Ab0} \alpha^{-\pi}}{k_{Af}A_{f0}} = \frac{K_{d0}}{\alpha^{\pi} A_{f0}}$. This expression shows that with increase in $K_{d0}$, the velocity of the wavepacket decreases.

\section{Parameters}
\begin{table}[htpb]
\centering
\begin{tabular}{||m{1cm}|m{1.5cm}||m{1cm}|m{1.5cm}||m{0.75cm}|m{1cm}||}
    \hline
     $k_0$ & 2.5 & $\gamma$ & 0.5 & $\omega$ & $10^{-1}$\\
     \hline
     $k_1$ & 0$\cdot 10^{-2}$ & $\gamma_1$ & $10^{-2}$ & $K_d$ & $10^{-1}$ \\
     \hline
     $k_{dp}$ & 2$\cdot 10^{-1}$ & $\gamma_{dp}$ & $5 \cdot 10^{-2}$ & $\omega_1$ & 1 \\
     \hline
     $k_{Ab,0}$ & $10^3$ & $K_{d0}$ & 1-11 & $K_{d1}$ & $10^{-1}$ \\
     \hline
     $\alpha$ & 10 & $A_t$ & 0.1 & $\epsilon_{seq}$ & 0 \\
     \hline
\end{tabular}
\caption{Parameter Values for comparing Analytical prediction and Numerical Simulation in the $k_1 = 0$ case, i.e. when ultrasensitivity is absent.}
\label{ParameterValues_new}
\end{table}

\begin{table}[htpb]
    \centering
    \begin{tabular}{||m{1cm}|m{1.5cm}||m{1cm}|m{1.5cm}||m{0.75cm}|m{1cm}||}
        \hline
         $k_0$ & 2.5 & $\gamma$ & 0.08 & $\omega$ & $10^{-1}$\\
         \hline
         $k_1$ & 1-5$\cdot 10^{-2}$ & $\gamma_1$ & $10^{-2}$ & $K_d$ & $10^{-2}$ \\
         \hline
         $k_{dp}$ & 5$\cdot 10^{-2}$ & $\gamma_{dp}$ & $5 \cdot 10^{-2}$ & $\omega_1$ & $10^{-2}$ \\
         \hline
         $k_{Ab,0}$ & $10^3$ & $K_{d0}$ & 1-11 & $K_{d1}$ & $10^{-2}$ \\
         \hline
         $\alpha$ & 10 & $A_t$ & 0.1 & $\epsilon_{seq}$ & 0.1 \\
         \hline
    \end{tabular}
    \caption{Parameter Values for Numerical simulations in the $k_1 \neq 0$ case, i.e. when ultrasensitivity is present.}
    \label{ParameterValues}
\end{table}

All numerical simulations were performed using ODE15s function of MATLAB. In all our simulations, the value of N was 100, i.e. there were 101 states of each type, $P_1$, $P_2$ and $P_3$.

\end{document}